\documentclass[%
 reprint,
 superscriptaddress,
 amsmath,amssymb,
 aps,
 prx,
]{revtex4-2}

\usepackage{graphicx}
\usepackage{dcolumn}
\usepackage{bm}
\usepackage{hyperref}
\usepackage[mathlines]{lineno}%

\begin{document}

\preprint{APS/123-QED}

\title{Exploring the tropical Pacific manifold in models and observations}

\author{Fabrizio Falasca}
\email{fabri.falasca@nyu.edu}
\affiliation{Courant Institute of Mathematical Sciences, New York University, New York, NY, USA
}
\affiliation{School of Earth and Atmospheric Sciences, Georgia Institute of Technology, Atlanta, GA, USA
}
\author{Annalisa Bracco}%
\affiliation{School of Earth and Atmospheric Sciences, Georgia Institute of Technology, Atlanta, GA, USA
}

\date{\today}

\begin{abstract}
The threat of global warming and the demand for reliable climate predictions pose a formidable challenge being the climate system multiscale, high-dimensional and nonlinear. Spatiotemporal recurrences of the system hint to the presence of a low-dimensional manifold containing the high-dimensional climate trajectory that could make the problem more tractable. Here we argue that reproducing the geometrical and topological properties of the low-dimensional attractor should be a key target for models used in climate projections. In doing so, we propose a general data-driven framework to characterize the climate attractor and showcase it in the tropical Pacific ocean using a reanalysis as observational proxy and two state-of-the-art models. The analysis spans four variables simultaneously over the periods 1979-2019 and 2060-2100. At each time $t$, the system can be uniquely described by a state space vector parameterized by N variables and their spatial variability.  The dynamics is confined on a manifold with dimension lower than the full state space that we characterize through manifold learning algorithms, both linear and nonlinear. Nonlinear algorithms describe the attractor through fewer components than linear ones by considering its curved geometry, allowing for visualizing the high-dimensional dynamics through low-dimensional projections.  The local geometry and local stability of the high-dimensional, multi-variable climate attractor are quantified through the local dimension and persistence metrics. Model biases that hamper climate predictability are identified and found to be similar in the multivariate attractor of the two models during the historical period while diverging under the warming scenario considered. Finally, the relationships between different sub-spaces (univariate fields), and therefore among climate variables, are evaluated. The proposed framework provides a comprehensive, physically based, test for assessing climate feedbacks and opens new avenues for improving their model representation.  
\end{abstract}

\maketitle


\section{Introduction} \label{sec:level1}

A comprehensive analysis of climate variability should account for the multivariate and nonlinear dependencies intrinsic to the system. Quantifying these dependencies is an urgent challenge in climate research \cite{Knutti}. Dynamical system theory offers a pathway to this end, so far successfully applied to high-dimensional turbulent systems \cite{Hopf, Predrag}. At each time step $t$, the state of a spatiotemporal chaotic system can be viewed as a point in an infinite dimensional state space, parametrized by multiple variables and their spatial dependency. In the case of dissipative chaotic systems, this infinitely dimensional dynamics is confined on a finite dimensional object, commonly known as “inertial manifold” or “attractor” \cite{foias, DingPredrag}. Spatiotemporal chaos can then  be  seen  as  a  walk  on this lower dimensional inertial manifold \cite{Hopf,Predrag}.
Studies of this kind have contributed a quantitative understanding of moderate Reynolds-number turbulence \cite{Predrag,Gibson,Clockwork,Christiansen1997} and similar approaches have been applied to a variety of fields, from computational neuroscience \cite{Solla,REMINGTON2018938,SOHN2019934,jazayeri2021interpreting,Churchland} to artificial neural networks (ANN) \cite{Cohen, stephenson2021on} and biophysics \cite{Ahamed,Stephens}. Each one of these fields present the challenge of identifying low-dimensional manifolds embedded in very high-dimensional, noisy data \cite{Predrag,DingPredrag}. Computer  scientists  have  been long interested in identifying low dimensional manifolds from data as dimensionality reduction tools, a problem commonly known in the literature as “manifold learning” (e.g., \cite{saul}).  The Earth's climate system is indeed a high dimensional, dissipative dynamical system and its dynamics is expected to be confined to a manifold with lower dimension than the full state space \cite{GhilLucarini}. A recent application of manifold learning in climate science has shown the potential of this idea by identifying the multistability properties of an intermediate complexity climate model \cite{Margazoglou}.

In this work, we propose a dynamical systems framework to investigate and compare spatiotemporal climate variability in observations and state-of-the-art climate. The climate system is highly nonlinear and climate models are far from perfect \cite{lenny}. As a result, climate projections require ensemble simulations to test for sensitivity to initial conditions, and different models \cite{lenny,Frigg} or strategies \cite{Katarzyna} to cope with model errors. Intermodel comparisons, such as the Coupled Model Intercomparison Project \cite{gmd2016}, generate petabytes of data. Mining and quantifying sources of biases, limitations and ambiguities among these data is fundamental when communicating results to other scientists and policy-makers \cite{lennyUncertainty,Eyring}. In  this context, the dynamical system view that we propose brings several advantages: it allows to comprehensively study the system accounting for, and quantifying, multivariable dependencies, to assess both mean values and instantaneous properties of each state  space  point, and to curtail the need for large ensembles in the evaluation of climate model biases and future trajectories. The characterization of the manifold dimensionality of a modeled climate system is indeed largely independent of the realization or ensemble member considered if sufficiently well sampled. For a given climate or time period, the modeled attractor does not change.\\

The focus of this work is the Pacific Ocean between 20$^o$N and 20$^o$S, where the  El  Ni\~{n}o  Southern  Oscillation  (ENSO) is the main mode of climate variability. ENSO is an oscillatory mode  driving,  with  its  warm  and  cold  phases,  El  Ni\~{n}o and La Ni\~{n}a, the  most  dramatic  year-to-year  variation of the Earth’s climate system \cite{ENSOcomplexity}. ENSO affects rainfall patterns, tropical cyclogenesis  and the likelihood of droughts and floods, and freshwater availability. ENSO impacts also food security, with cascading effects on health, water, sanitation, education, and overall increased mortality \cite{ENSOcomplexity,Bronnimann,YuEnso,HeEnso}. In light of its great societal relevance, the tropical Pacific is a much studied region, and therefore a convenient, well known, test-case. Our approach, however, is general and could be applied anywhere else at both global and regional scales. Model evaluations for ENSO dynamics commonly employ traditional methodologies, such as power spectra and standard deviations \cite{ecEarth,cm4,ipsl}, or more advanced methods such as percolation theory \cite{percolation} and complex networks \cite{Ilias2, Falasca}. Current analyses, however, do not routinely adopt a multivariate approach and quantify only average dynamical properties. Furthermore, the high dimensionality of the system is very seldom considered, and the analysis is often limited to one component, associated to the Ni\~{n}o3.4 index or the first principal component \cite{PCA}.\\

Here, we consider a subset of four key variables relevant to the tropical Pacific dynamics: surface temperature, zonal and meridional surface velocities and outgoing longwave radiation, and perform our analysis on the ERA5 reanalysis \cite{ERA5} and two state-of-the-art climate models from the Coupled Model Intercomparison Project – phase 6 (CMIP6) \cite{cmip6}, the Max Planck Institute (MPI) and EC-Earth models \cite{mpiModel,ecEarth}.
We analyze daily data and their anomalies over two 40 years periods, 1979-2019 and 2060-2100. The two models were chosen because of their output availability and significant differences in their parameterizations, resolutions and performance \cite{Fasullo}. We visualize the high dimensional inertial manifold using both a linear, Principal Component Analysis (PCA) \cite{PCA}, a nonlinear, Isomap \cite{ISOMAP}, dimensionality reduction methods. Isomap has been chosen among many others (e.g. \cite{rockPCA,Belkin,mcinnes2020umap,tsne,lle} - see \cite{chung2021neural} for a brief review) for its simplicity. Using a nonlinear method has the advantage of  identifying the curved manifold itself (intrinsic dimension), rather than the embedding space found by PCA, as discussed in \cite{chung2021neural,jazayeri2021interpreting}.\\

The  time series that we obtain describe the time-evolution of modes of climate variability, accounting for  more than one variable, and potentially could include all or most model variables; the linear regression of these time series onto each original spatiotemporal field defines the spatial signature of the modes. Most importantly, by comparing  projections in the multivariate and univariate representation we can highlight the role of each variable in the overall system’s dynamics and its relation with the other fields.\\

We then estimate the local properties and stability of the climate attractor of the tropical Pacific in the reanalysis and the models. For a given climate state $\zeta$ (i.e., configuration of the Pacific at an instant in time), we  evaluate the geometrical properties of the attractor in terms of its local dimension metric, $d(\zeta)$ \cite{FarandaSciRep,FarandaPaleo} and its stability, $\theta(\zeta)$ through the inverse of the average persistence of the trajectory around $\zeta$ \cite{Moloney,Caby}. While $d$ roughly quantifies the number of directions the system can evolve from/into, and therefore the number of degrees of freedom required to describe it at that point in time, the persistence quantifies the ``stickiness'' 
of the trajectory around each neighborhood in state space, and therefore how predictable the future evolution of that state is, or its predictability potential. These concepts have been first introduced in \cite{Freitas, LucariniBook} and have been applied so far to univariate fields to explore atmospheric weather regimes \cite{FarandaSciRep,FarandaPaleo,MessoriJets,DeLuca,DeLucaQJRMS,Rodrigues,MessoriGRL,MessoriHeatWaves,Hochman,GhilLucarini}.\\ Finally, we explore sub-spaces of the original state space by evaluating local dimension and persistence of each univariate field. While these properties are expected to differ between models and observations  being  the modelled manifold generally lower dimensional by construction \cite{Dubrulle}, a reliable model should capture the  observed relationships among variables to properly represent climate feedbacks, and therefore the relative scaling of such metrics.\\

The data analyzed are described in section \ref{sec:data}. The analysis framework is presented in section \ref{sec:method}, followed by the results (section \ref{sec:results}). A discussion of our results and their implications, and future avenues of research concludes this work.

\section{Data} \label{sec:data}
The ERA5 reanalysis \cite{ERA5} is the observational-based dataset adopted. Produced by the European Centre for Medium-Range Weather Forecasts (ECMWF), ERA5 data are available at a spatial resolution up to 31 km globally. We consider two models, the Max Planck Institute  MPI-ESM1.2-HR  and  EC-Earth3-HR  (MPI and EC-Earth hereafter) \cite{MPIHR,ecEarth}. The resolution of their atmosphere and ocean components is  $\sim 100$ km and 0.4$^o$ for MPI , and $\sim 40$ km and 0.25$^o$ for EC-Earth. Both datasets are part of the Coupled Model Intercomparison Project – phase 6 (CMIP6) \cite{cmip6} catalog. Their performance according to the scoring analysis by \cite{Fasullo} is in the top five (EC-Earth) and middle range (MPI) among the 37 model configuration tested over the historical period.

The analysis spans the periods 1979-2019 and 2060-2100 at daily frequency. Given that the CMIP6 historical integrations end on December 2014,  we concatenated 4 years from the SSP585 scenario, the ``worst-case scenario'' with the highest radiative forcing (up to 8.5 $W/m^{2}$ in 2100). We further analyzed the SSP585 outputs for the last four decades of the XXI century, 2060-2100. 

For each model we analyze 4 members, randomly chosen,  in the historical period and 4 for EC-Earth and 2 for MPI (only two are available with daily outputs) in the future.  Our discussion, however, focus on one member, the first in the respective ensembles, in most of our presentation (CMIP6-label: r1i1p1f1 for both models). In the last section of this work (Section \ref{sec: internal_variability}), we show that the attractor characteristics, as quantified by the local dimension and persistence metrics, do not change as function of the ensemble member considered. In fact, chaotic trajectories of the same dynamical system are still bounded to live on the same manifold and manifold properties should not depend on the ensemble member, provided that we sample such object well enough.

The domain of interest is the tropical Pacific in the latitude-longitude range [20$^o$S-20$^o$N, 120$^o$E-70$^o$W]. All dataset are remapped on a coarser grid with resolution of 1$^o$ in latitude by 1.5$^o$  in longitude; a reasonable step as we are interested in large scale dynamics. A higher latitudinal resolution is chosen to ensure we are resolving the Rossby wave field \cite{held}.\\
The variables considered are surface temperature (T), zonal and meridional velocities (u,v) at the surface and outgoing longwave radiation (OLR). The temperature field is a driver of variability in the Pacific ocean; the horizontal velocity vector field quantifies the dynamical response of the system; and the outgoing longwave radiation is a proxy for cloud variability which  is key during ENSO. For reproducibility purposes, the variables chosen are: temperature at 2m (label: t2m), zonal and meridional velocity at 10m (label: u10 and v10) and the top net thermal radiation (label ttr and equal to the negative of OLR) in ERA5, and temperature and velocities at the surface (labels: tas, uas and vas) and the outgoing longwave flux at the top of the atmosphere (label: rlut) in the models.

\textbf{State space embedding.} Given the choice of fields, at each time step $t$, the tropical Pacific climate is uniquely described by a state space vector defined by  $\mathbf{X} = [\text{T}(x,y), \text{u}(x,y), \text{v}(x,y), \text{OLR}(x,y)](t) \in \mathbb{R}^{\textit{T,N}}$. $T$ is the length of each time series and given that we consider separately two forty-years long periods at daily temporal resolution, $T = 14,975$ days. $N$ is the dimensionality of the state space and for the spatial resolution and variables considered, $N=17,092$. 

\section{Earth's Climate as a dynamical system} \label{sec:method}

An important novelty of our approach is that we study the evolution of the highly dimensional climate system focusing on multiple variables simultaneously \cite{Predrag}.\\ Let us consider a spatiotemporal  climate system described by $m$ fields $\mathbf{Y}_{i}$, $i = 1,2,...,m$, embedded in a grid of size $g$ and spanning a time interval $T$. For each field $\mathbf{Y}_{i}$, first we weight all time series by the cosine of their latitude. Each field is then standardized to zero mean and unit variance. If velocity fields are included, it is crucial to standardize the velocity vector and not each single component separately.\\
At each time step $t$, the system is uniquely described by a state space vector $\mathbf{X} = [\mathbf{Y}_{1},\mathbf{Y}_{2},...,\mathbf{Y}_{m}](t) \in \mathbb{R}^{\textit{T,N}}$ of dimensionality $N = m ~\text{x} ~g$. In this formulation, a single- point trajectory in state space describes the climate system evolution. This trajectory spans a manifold with lower dimension than the full state space because the system is dissipative \cite{GhilLucarini}.\\
Several methodologies can then be applied to investigate such high-dimensional dynamics, as briefly summarized below.

\subsection{Manifold learning} \label{sec:manifold_learning}

The identification of modes of variability in climate science often relies on Principal Component Analysis (PCA) \cite{PCA}, commonly referred to as Empirical Orthogonal Functions (EOF) and its variations, such as rotated-EOF (R-EOF) \cite{Kawamura} and Extended EOF \cite{weareEOF}. In most climate applications, nonlinear components are neglected (for an exception see e.g. \cite{rockPCA}), and variables are investigated one at the time. However, the climate system is comprised by many interacting, covarying variables and a multivariate approach represents a more rigorous way of quantifying its dynamics. State space embedding for the climate system as a whole is an ill-posed problem, due to the very large number of variables spanning physical, chemical and biological processes, and not all governing equations are known. For specific problems, on the other hand, a subset of key variables considered together, for example those for which we have a good observational record of sort, can offer an in-depth dynamical understanding of the system. To visualize the geometry of the underlying manifold the linear, Principal Component Analysis \cite{PCA} may still be adopted, or we can rely on nonlinear methods, such as Isomap \cite{ISOMAP}.\\

\paragraph{Principal Component Analysis.} \label{sec:pca}

Principal Component Analysis (PCA) \cite{PCA}, or Empirical Orthogonal Function (EOF)  analysis \cite{storchzwiers}, is a linear modal decomposition method. Given a spatiotemporal dataset $\mathbf{X} \in \mathbb{R}^{\textit{T,N}}$ with $N$ time series each of length $T$, PCA identifies the underlying manifold by fitting hyperplanes in the directions that contain most of the variance. This is achieved by computing the Gram matrix as $\mathbf{G} = \frac{1}{\textit{T}-1}\mathbf{X}\mathbf{X}^{\text{T}} \in \mathbb{R}^{\textit{T,T}}$ \cite{ROCK-PCA}. The $T$ eigenvectors $\mathbf{U}\in\mathbb{R}^{\textit{T,T}}$ of the Gram matrix $\mathbf{G}$ are the Principal Components (PCs) of the dataset. Alternatively, it is possible to eigen-decompose the covariance matrix $\mathbf{C} = \frac{1}{\textit{N}-1}\mathbf{X}^{\text{T}}\mathbf{X} \in \mathbb{R}^{\textit{N,N}}$. In this case the eigenvectors $\mathbf{V}\in\mathbb{R}^{\textit{N,N}}$ of $\mathbf{C}$ are spatial patterns and the projection of  $\mathbf{V}$ onto $\mathbf{X}$ describes their temporal variability. The decomposition of the Gram matrix returns the same solution of the covariance matrix up to $T$ eigenvectors \cite{Bishop,ROCK-PCA}.  A third, equivalent alternative, is to eigen-decompose the Euclidean distance matrix containing the (Euclidean) distances between each point in $\mathbf{X}$ with the multidimensional scaling (MDS) algorithm \cite{Borg,saul}.
Each $\text{U}_{i}$ explains a given variance based on the ratio of its correspondent eigenvalue and the sum of all eigenvalues, i.e. $\lambda_{i}/\sum_{j}\lambda_{j}$. To identify the correspondent spatial patterns,  is enough to linearly regress each (standardized) PC on the dataset as $\frac{1}{\textit{T}-1}\mathbf{U}^{\text{T}}\mathbf{X}$. The low-dimensional projection found by PCA preserves the variances as measured in the high-dimensional input $\mathbf{X}$ \cite{ISOMAP}.\\

\paragraph{Isometric feature mapping (Isomap).} \label{sec:isomap}
If the manifold is nonlinear Euclidean distances cannot capture the intrinsic distances along the manifold. Isomap is one of the available tools to address this problem (see \cite{chung2021neural}) and has been introduced to climate science by \cite{Ross} and \cite{Hannachi} for univariate fields. Isomap adds a key step to the MDS algorithm. Given a dataset $\mathbf{X} \in \mathbb{R}^{\textit{T,N}}$, with $N$ time series each of length $T$ and  centered to zero mean,  Isomap first identifies the $K$-nearest neighbors of each point $i$ in the manifold; then it computes the \textit{geodesic} distances $\delta_{i,j}$ between each couple of points $i$ and $j$ by assuming that the manifold is \textit{locally} flat in a radius of $K$ points (see also App. A). The geodesic distance matrix $\mathbf{D}_{g}$ is finally used as input to the MDS algorithm \cite{Borg}. Given $\mathbf{D}_{g}$, the double centered distance matrix is computed as $\mathbf{A} = - \frac{1}{2} \mathbf{J}\mathbf{D}_{g}\mathbf{J}$, with $\mathbf{J} = \mathbf{I}_{T}-\frac{1}{T}\mathbf{e}\mathbf{e}^{\text{T}}$ where $\mathbf{I}_{T}$ is the identity matrix of order $T$ and $\mathbf{e}$ is a vector of length $T$ containing all ones.\\
The dimensionality of the dataset is then reduced by finding the eigenvectors of the double centered matrix $\mathbf{A}$  (i.e., solving $\mathbf{A} = \mathbf{U}\mathbf{\Lambda}\mathbf{U}^{\text{T}}$). Each component $\text{V}_{i}$ is  obtained by weighting the $i$-th eigenvector $\text{U}_{i}$ by the square root of its correspondent eigenvalue $\sqrt{\lambda_{i}}$, as $\text{V}_{i} = \text{U}_{i}\sqrt{\lambda_{i}}$ \cite{ISOMAP}. Similarly to PCA, the associated spatial patterns can be retrieved by linearly regressing the (standardized) Isomap components on the original dataset as $\frac{1}{\textit{T}-1} \mathbf{V}^{\text{T}} \mathbf{X}$.\\

A downside of using nonlinear dimensionality reduction methods is that the explained variance cannot be directly estimated. The so-called \textit{residual} variance has been proposed as valid alternative. It is defined as $1 - R(D_{M}, D_{Y})^{2}$, where $D_{M}$ is the algorithm's estimate of manifold distances (i.e, Euclidean distance matrix for PCA and geodesic matrix for Isomap), $D_{Y}$ is the matrix of Euclidean distances in the low dimensional embedding computed by each algorithm and $R$ is the Pearson correlation coefficient among all entries of $D_{M}$ and $D_{Y}$ \citep{ISOMAP}.\\

As mentioned, Isomap relies on the parameter $K$. In our subsequent analysis we set $K = 10$, thus assuming that the manifold is \textit{locally} flat within a radius of 10 days, which is reasonable given the spatial resolution considered and the length of our time-series.

\subsection{Dynamical systems metrics: local dimension and persistence} \label{sec:quantitative}

Quantifying the dimensionality of the inertial manifold allows to estimate the \textit{effective} degree of freedoms of the dynamical system being investigated, and therefore its complexity \cite{theiler}. This is a difficult problem for spatiotemporal chaotic systems and an \textit{exact} estimation of such quantity may require knowing the equation of motions, as done in \citet{DingPredrag} for the Kuramoto-Sivashinsky system. For very high-dimensional, noisy datasets, data-driven methods have shown limitations \cite{saraDimensionality} and depending on the problem at hand, certain approaches may be more useful than others.\\

In the case of climate variability, recent advances at the interface of dynamical system and extreme value theory have opened the possibility to estimate the dimensionality of underlying attractors \cite{LucariniBook}. For a given dynamical system, the probability of recurrence of a state $\zeta$ obeys a generalized Pareto distribution. Locally, such distribution scales with a parameter, shown to be equal to the local dimension $d(\zeta)$. The attractor dimensionality (i.e., Hausdorff dimension) can be retrieved by averaging all $d$ \cite{locDim,pons}. Furthermore, for a climate system it is useful to quantify the tendency of the trajectory to persist in a neighbor of the state space, because this tendency is directly linked to the predictability of that state. The greater the persistence, the higher the predictability. This property can be quantified by introducing the so-called extremal index \citep{Moloney}. The two metrics, local dimension and persistency, are powerful tools to explore high dimensional dynamics and have been useful to characterize univariate atmospheric fields in several recent studies (e.g., \cite{FarandaSciRep,DeLucaQJRMS,MessoriGRL,Hochman} among others).\\

Here we briefly present these tools and refer the reader to \cite{LucariniBook} for details and rigorous proofs.\\

We consider the high-dimensional trajectory $\mathbf{X}(t) \in \mathbb{R}^{\textit{T,N}}$ and for each state $\zeta = \mathbf{X}(\tau)$, with $\tau \in [1,T]$, we define the observable as
\begin{align}
g(\mathbf{X}(t),\zeta) = - \log(\delta(\mathbf{X}(t),\zeta))~.
\label{eq:observable}
\end{align}
Here $\delta(x,y)$ represents the Euclidean distance between two vectors $x$ and $y$ and the logarithm further increases the discrimination between close recurrences \cite{LucariniBook,FarandaSciRep}. The minus sign turns minima into maxima for practical convenience, thus the time series $g$ is large when $\mathbf{X}(t)$ is similar to $\zeta$.\\
We then define a threshold $s(q,\zeta)$ as the $q$th quantile for each $g(\mathbf{X}(t),\zeta)$ and adopt $q = 0.98$, as in \cite{FarandaSciRep}. The sensitivity to the choice of q value is addressed in Appendix \ref{app:param_q}. The points in $g(\mathbf{X}(t),\zeta)$ that exceed the threshold $s(q,\zeta)$ represents the Poincar\'{e} recurrences of $\zeta$ and are here referred to as $u(\zeta)$ \cite{LucariniBook}. Finding recurrences of a state $\zeta$ in a neighborhood of radius $r$ is therefore equivalent to find exceedances of $g(\mathbf{X}(t),\zeta)$ over a threshold $s$, and we refer to a neighborhood of a state $\zeta$ as $\Gamma_{q}(\zeta)$. 
The Freitas-Freitas-Todd theorem \cite{Freitas}, modified by Lucarini et al. (2016) \cite{LucariniBook}, states that the probability $P(u,\zeta)$ that the dynamics $\mathbf{X}(t)$ returns in a neighborhood $\Gamma_{q}(\zeta)$ converges to a Generalized Pareto Distribution \cite{Pickands}:
\begin{align}
P(u,\zeta) \simeq \exp{[-\theta(\zeta) \frac{u(\zeta)}{\sigma(\zeta)}]} ~;
\label{eq:pareto}
\end{align}
where $\sigma(\zeta)$ and $\theta(\zeta)$, the so-called \textit{extremal index} \cite{Moloney,Leadbetter}, are parameters of the distribution.\\

The \textit{local dimension} $d(\zeta)$ can be computed as $d(\zeta) = \frac{1}{\sigma(\zeta)}$ \cite{LucariniBook}. It relates to the density of state space points in a neighborhood $\Gamma_{q}(\zeta)$ and roughly quantifies the number of directions the system can evolve from/into.  $d(\zeta)$  is therefore linked to the intrinsic \textit{local} predictability of $\zeta$ \cite{FarandaSciRep} (i.e., the lower is $d(\zeta)$ , the larger is the predictability of that state). \cite{LucariniBook} suggests that the attractor dimension can be computed as the average over all local dimensions $D = \langle d(\zeta) \rangle$ and \cite{FarandaSciRep} has shown that this is indeed the case for the Lorenz system \cite{Lorenz}.\\

The \textit{persistence} of the trajectory $\mathbf{X}(t)$ in a neighborhood $\Gamma_{q}(\zeta)$ is quantified by the extremal index $\theta(\zeta)$ \cite{Moloney,Suveges,Ferro,Weissman}. Here $\theta$ is computed using the methodology proposed by S\"{u}veges (2007) \cite{Suveges}. Intuitively, $\theta \in [0,1]$ is linked to the inverse of the mean residence time of $\mathbf{X}(t)$ in $\Gamma_{q}(\zeta)$, with low values implying higher persistence in the neighborhood. Higher persistence around a state $\zeta$ quantifies the tendency of the trajectory to stick in its neighborhood, therefore increasing the potential predictability around $\zeta$. The value of $\theta(\zeta)$ should be divided by the time step $\Delta t$  (i.e., $\theta(\zeta)/\Delta t$), but in this paper we are in the trivial situation of $\Delta t$ = 1 day. For the Lorenz system, lower $\theta$ are found in the neighborhoods of the 3 \textit{unstable} fixed points \citep{FarandaSciRep}.\\

The local dimension and persistence metrics are adopted after removing the seasonality and the trend. This allows to focus on stationary time series.

\section{Nonlinear and multivariate dimensionality reduction} \label{sec:results}

\subsection{A first test: Mean State and Seasonal Cycle} \label{sec:raw_data}

\begin{figure*}[tbhp]
\centering
\includegraphics[width=0.8\linewidth]{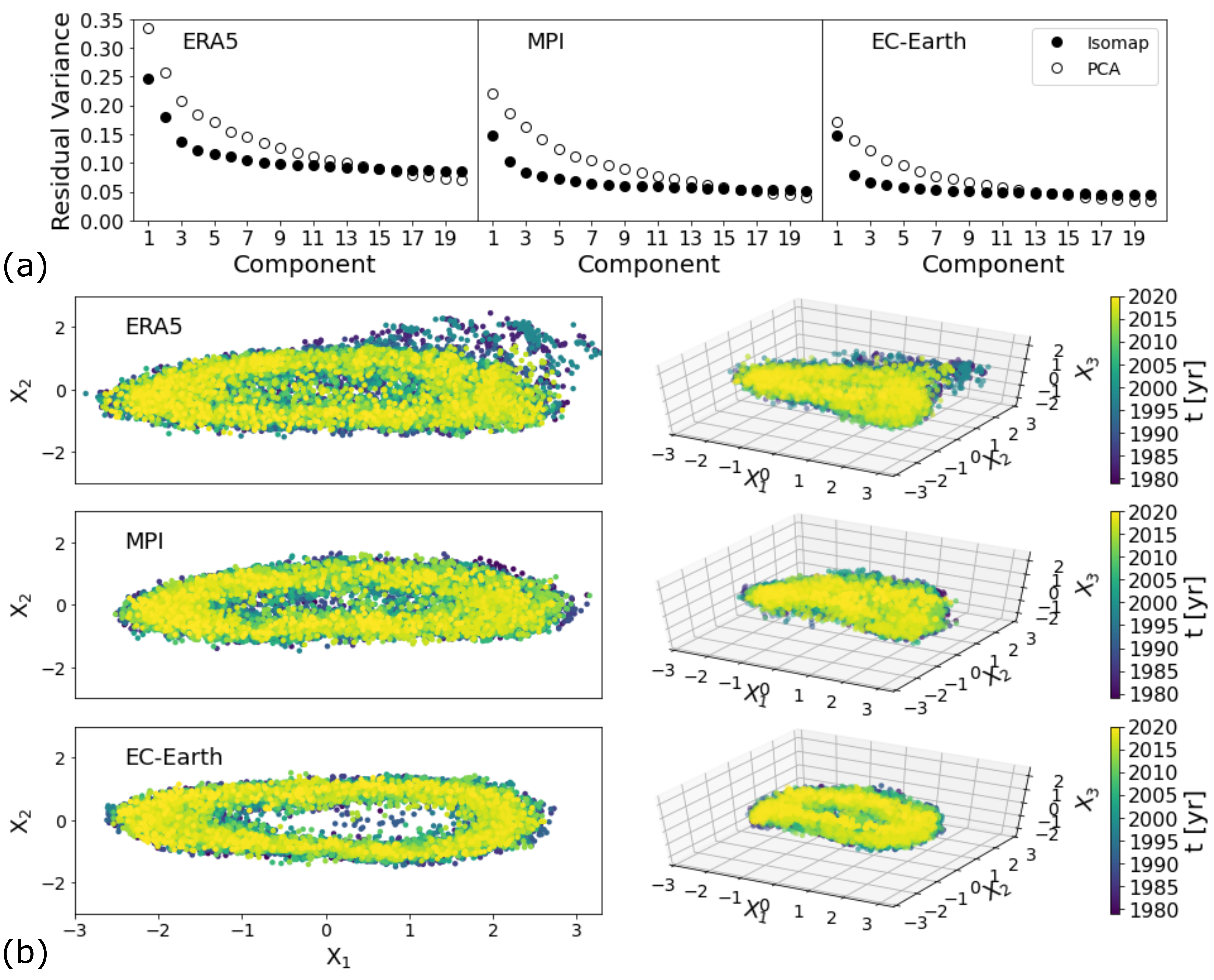}
\caption{Dimensionality reduction of the tropical Pacific climate retaining seasonality and trends. Panel (a): residual variance of PCA and Isomap for ERA5 and the two CMIP6 models (MPI and EC-Earth). Panel (b): low dimensional, Isomap projections (first 2 or 3 components, respectively) of the three datasets; Isomap components are indicated as $X_{1}, X_{2}, X_{3}$ respectively. Each point encodes the multivariate state of the tropical Pacific ocean at a given day from January 1979 to December 2019. Projections have been standardized to unit variance.}
\label{fig:state_space_raw_data}
\end{figure*}

As a proof of concept, we consider the state space evolution in the ERA5, MPI and EC-Earth in the tropical Pacific retaining trends and seasonal cycle and considering all four fields mentioned. The residual variance after applying the PCA and Isomap algorithms (see Methods) on the embedded state space vector $\mathbf{X} = [\text{T}(x,y), \text{u}(x,y), \text{v}(x,y), \text{OLR}(x,y)](t)$  is shown in Figure \ref{fig:state_space_raw_data}a. In both observations and models, the residual variance of the first three components is lower in Isomap; additionally, Isomap saturates faster than PCA. In other words, the 17,092-dimensional trajectory of the tropical Pacific domain lives on a low dimensional object, which is non- linear, as verified by comparing Euclidean and geodesic distances for the 3 datasets (see App. A, Figure \ref{fig:distances}). 
The models share strong similarities in their average Euclidean distance but they differ from the reanalysis in their geodesic components.\\
Comparing ERA5 and models, the residual variance of
the first PCA and Isomap components is higher in the reanalysis, and saturates to a higher value. In EC-Earth, on the other hand, PC1 and Isomap-1 explain  a nearly identical amount of variance.

Focusing on the low dimensional Isomap projections, each point in Figure \ref{fig:state_space_raw_data}b represents the state of the Pacific system $\mathbf{X} = [\text{T}(x,y), \text{u}(x,y), \text{v}(x,y), \text{OLR}(x,y)](t)$ at a given day. Due to the inclusion of the seasonal cycle and its dominance on the overall variability, the dynamics lives on a torus, which is topologically similar among models and reanalysis. The ERA5 dynamics, however, deviates from the periodic trajectory in correspondence of  the  1982/83  and  1997/98 El Ni\~{n}o events.  No  clear deviation occurs in the models at any time, indicating that no ENSO event was capable of modifying the modeled seasonal cycle in the simulations considered   \cite{Maher}.\\
The correlation between Isomap-1 and PC1 is higher than 0.98 independently of the dataset, illustrating that the seasonal cycle variability is a close-to-linear process, as expected. Furthermore, the Pearson correlation across datasets is  higher than 0.95 in all cases, indicating that the temporal characteristic of the seasonal cycle are well captured by both models. Large correlations across datasets, however, do not imply similar spatial projections (see Methods), which also consider differences in signals' variance: large regional biases in the representation of the seasonal cycle are verified, as shown in App. B, Figure \ref{fig:seasonality}.

\subsection{Variability: ENSO} \label{sec:anomalies}

\begin{figure*}[tbhp]
\centering
\includegraphics[width=0.8\linewidth]{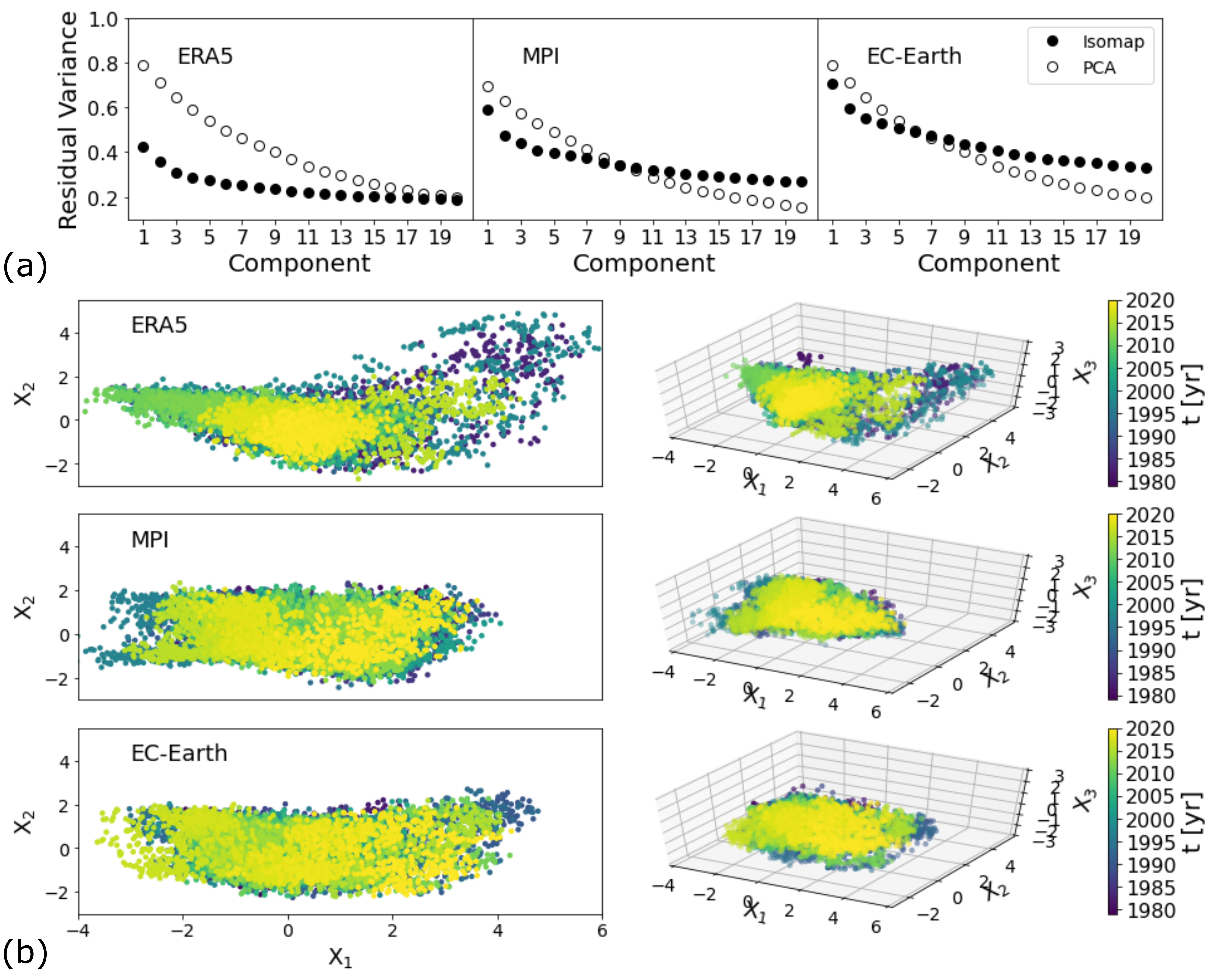}
\caption{Dimensionality reduction of the tropical Pacific Ocean anomalies (no linear trend and no seasonal cycle). Panel (a): residual variance of PCA and Isomap for observations (ERA5) and the two CMIP6 models (MPI and EC-Earth). Panel (b): low-dimensional, Isomap projections (first 2 or 3 components, respectively) of the three datasets; Isomap components are indicated as $X_{1}, X_{2}, X_{3}$ respectively. Each point encodes the multivariate state of the tropical Pacific ocean at a given day from January 1979 to December 2019. Projections have been standardized to unit variance.}
\label{fig:state_space_anomalies}
\end{figure*}

We next investigate the interannual variability of the tropical Pacific by repeating the analysis after de-trending and de-seasonalizing the data.  PCA and Isomap are again applied to the embedded state space vector $\mathbf{X} = [\text{T}(x,y), \text{u}(x,y), \text{v}(x,y), \text{OLR}(x,y)](t)$  and the residual variances are shown in Figure \ref{fig:state_space_anomalies}a.  For ERA5, the first Isomap component explains considerably more variance than the first PC (residual variances are 0.42 and 0.78 for Isomap-1 and PC1, respectively), and the first three Isomap components capture $\sim70\%$ of the total variance. Higher Isomap components show a faster saturation than PCs in residual variance but never fully saturate, underlying the high-dimensionality of the manifold in all three datasets. The PCs residual variance behaves similarly among models and reanalysis, while the Isomap residual variance in ERA5 differs from the models, suggesting that MPI and EC-Earth struggle in capturing the nonlinear topological characteristics of the reanalysis. In App. A, Figure \ref{fig:distances} we prove that also the manifold of the anomalies is nonlinear in all cases.\\
Differences among datasets emerge in the low dimensional 2D and 3D Isomap projections of the state space trajectory (Figure \ref{fig:state_space_anomalies}b). The strong 1982/83 and 1997/98 El Ni\~{n}os, followed by the 2015/16 event, are excursions away from the state space region usually explored by the tropical Pacific trajectory. Neither model replicates such behavior and the state space occupation suggests a structural difference between ERA5 and MPI/EC-Earth.   The residual variance explained by different components and the state space occupation are similar among models. This is verified despite the different ENSO characteristics quantified by the first Isomap and PC components and their power spectrum (see App. C, Figure \ref{fig:mode_1}). The spectral analysis identifies the largest, significant peak at $\sim 3.7$ year in ERA5 and EC-Earth, and at more than $\sim 8$ years in MPI.  In ERA5, PC1 and Isomap-1 projections are well correlated ($\text{r} = 0.9$) with largest discrepancies in correspondence of the 1982/83 and 1997/98 events. The peak, correspondent to the strong 2015/16 El Ni\~{n}o event is similarly identified in the PC1 and Isomap-1 components. Therefore, differently from the PC analysis, the Isomap projection implies that the 2015/16 event was of smaller amplitude and ``more linear'' in behavior compared to the 1982/83 and 1997/98 El Ni\~{n}os. In the models the correlations between PC1 and Isomap-1 projections are higher than in ERA5 ($\text{r} = 0.95$ for MPI and 0.94 for EC-Earth). \\

Spatial patterns are visualized by the regressions of Isomap-1 and shown in App. C, Figure \ref{fig:mode_1}. In ERA5, ENSO is characterized by a larger temperature variance on a narrow band in the equatorial eastern Pacific compared to the models, accompanied by a surface wind response consistent with a shift in the convective cell over the central to East Pacific and negative OLR anomalies (or more clouds) in the whole equatorial Pacific. In MPI the ENSO pattern is amplified in the central Pacific, as noted in \cite{MPIHR}, and the wind and cloud response is shifted to the western side of the basin. In EC-Earth the ENSO spatial signature in surface temperatures is closer to that of the reanalysis \citep{ecEarth}, but the atmospheric response remains biased similarly to MPI, with wind and OLR anomalies not extending sufficiently eastward and a visible double intertropical convergence zone (ITCZ) bias in cloud distribution.

By the end of the 21st century, ENSO dynamics change in the models in response to the increasing greenhouse gas concentrations of the SSP585 scenario.\\
The first components  - PC1 and Isomap-1 - of MPI, resemble the currently observed ENSO, with a spectral peak at $\sim 3.6$ years but no extreme El Ni\~{n}o events, while in EC-Earth positive and negative events recur periodically every 4 years (see App. C, Figure \ref{fig:mode_1_future}). ENSO amplitude increases slightly in MPI, with the largest changes in the OLR field, and substantially in EC-Earth, impacting temperature, zonal velocity  and OLR fields. In EC-Earth,  OLR changes the most,  with the ENSO-associated anomalies extending over the whole tropical Pacific ocean. Surface wind changes are concentrated in the zonal direction, and the Isomap-1 regression on the meridional wind velocity are nearly zero.

\subsection{Modes of variability: multivariate or univariate?} \label{sec:1_variable}
The framework proposed allows for analyzing simul- taneously many variables. Here we show the distinctive advantage of doing so.\\

\begin{figure}[tbhp]
\centering
\includegraphics[width=0.5\textwidth]{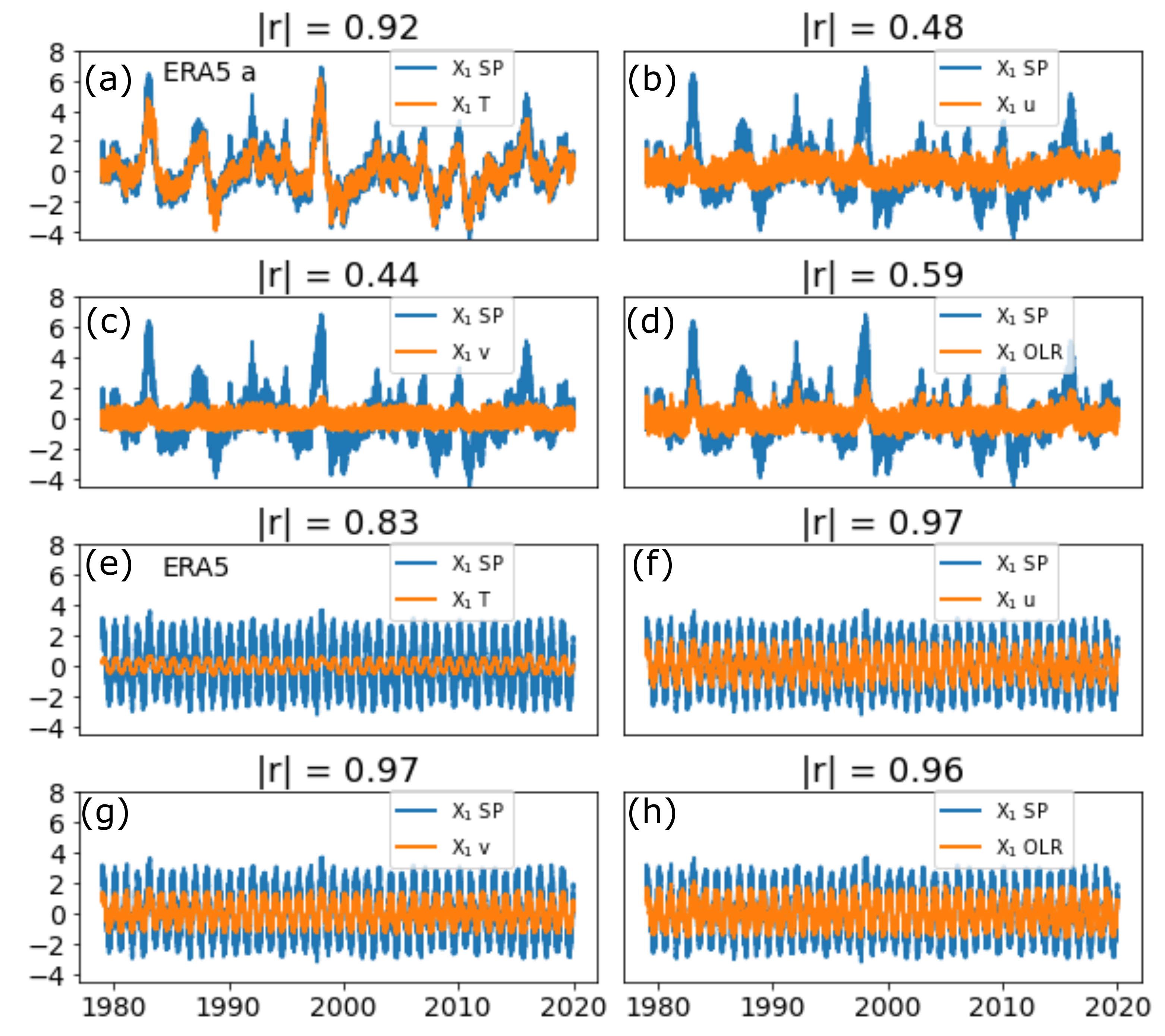}
\caption{
First Isomap ($X_{1}$) component for anomalies (top panel) and raw data (bottom panel) in ERA5. In each case components have been standardized so that the \textit{total} variance is equal to 1. Projections in the multivariate case are shown in blue and labelled as ``SP'' (state space). Projections for each, univariate field are shown in red. Atop of each plot is the correlation coefficient (in absolute value) between the projections in the multivariate and univariate cases.}
\label{fig:era5_multivariate_vs_univariate}
\end{figure}

In Figure \ref{fig:era5_multivariate_vs_univariate} we compare the first Isomap component obtained in the multivariate representation with the univariate case, applying Isomap separately to each field in ERA5. The corresponding figures for the models, looking only at the anomalies, in both current and future periods, can be found in Appendix \ref{app:1_variable}, Figures~\ref{fig:models_multivariate_vs_univariate_MPI} and~\ref{fig:models_multivariate_vs_univariate_EC_Earth}. Reducing the dimensionality in a multivariate representation corresponds to fitting axis in the direction that maximizes the overall variance. At times, the evolution of a single field reflects the largest fraction of the dataset’s variance, leading to small differences between the univariate and multivariate cases. This is the case for the tropical Pacific, where the variability is largely controlled by temperature anomalies.\\

In general, a multivariate approach helps identifying which variable contributes the most to the dynamics of interest. For example, when considering the seasonal cycle (Figure \ref{fig:seasonality}), the evolution on the torus in correspondence of large El Ni\~{n}o events (1982/83, 1997/98 and 2015/16) does not follow the evolution of temperature, as evident by comparing Isomap-1 in Figure \ref{fig:era5_multivariate_vs_univariate}(e) (see also Appendix \ref{app:1_variable}, Figure \ref{fig:multivariate_vs_univariate_era5_onlyT}). The multivariate representation largely ignores the temperature contribution in 2015/16. The seasonal cycle variance is indeed largely dominated by the velocity and OLR fields as shown in  Figure \ref{fig:era5_multivariate_vs_univariate}(e-h). In other words, the dimensionality reduction with or without embedding highlights the weighted (by variance) relative contribution of each variable to the system dynamics.\\

\begin{figure}[tbhp]
\centering
\includegraphics[width=1\linewidth]{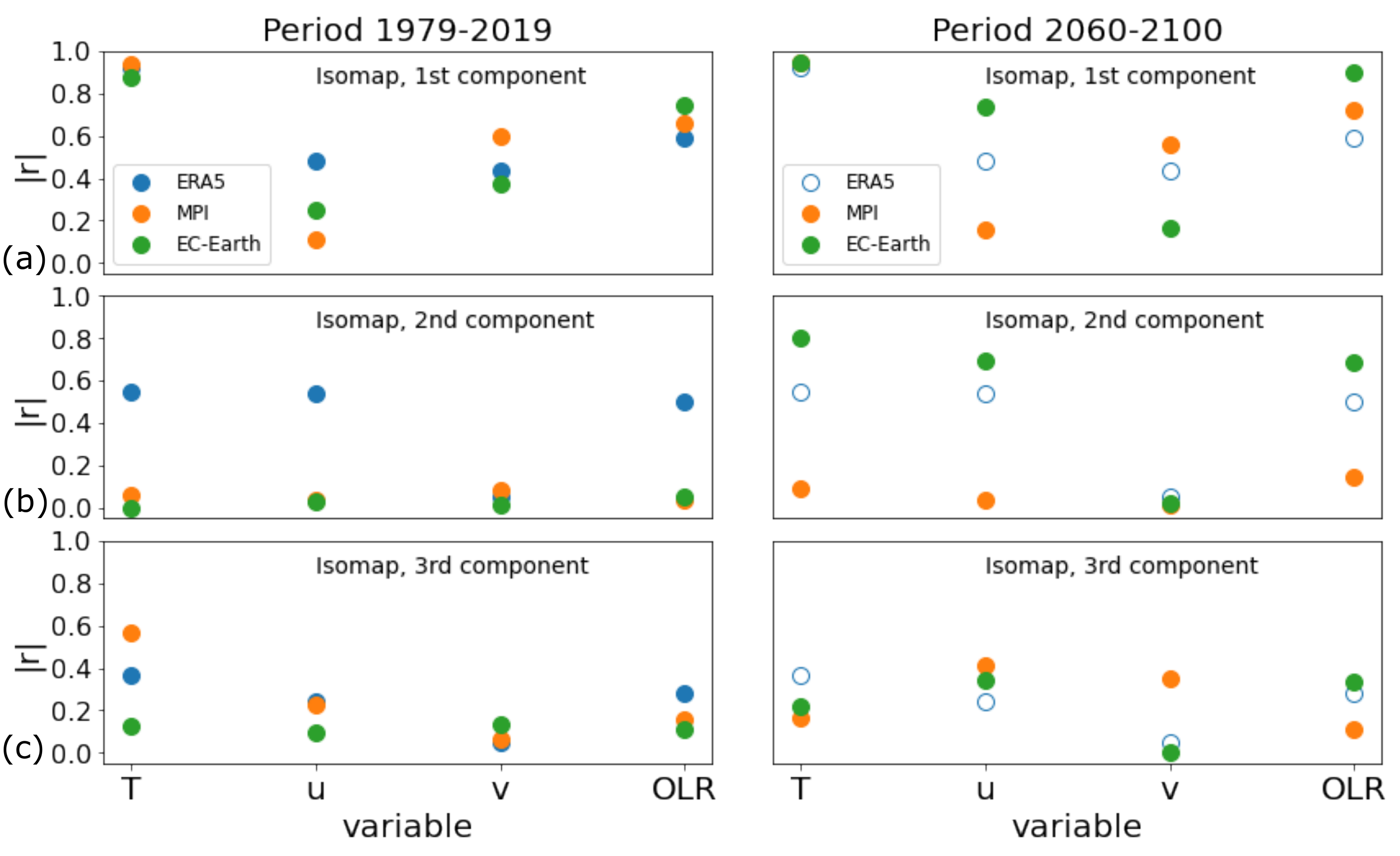}
\caption{Row (a): Correlations (in absolute value) between the first Isomap component of each variable (on the x-axis) and the multivariate case in ERA5 and models (signals in Figure \ref{fig:era5_multivariate_vs_univariate}(a-d) for ERA5 and Figures \ref{fig:models_multivariate_vs_univariate_MPI}, \ref{fig:models_multivariate_vs_univariate_EC_Earth} for MPI and EC-Earth respectively). In Row (b,d): same as row (a) but for the second and third Isomap components.}
\label{fig:correlations_multivariate_vs_univariate}
\end{figure}

Looking again at the anomalies, Figure \ref{fig:correlations_multivariate_vs_univariate} displays the correlations between the univariate and multivariate representations of the first three Isomap components in ERA5 and models in the two periods analyzed.\\ For the first component, the temperature field explains the largest part of the variance in both models and observations independently of the period analyzed, with correlations higher than 0.85 in all cases. Both models underestimate the relationship between the zonal wind u and the multivariate case, and MPI shows larger differences than EC-Earth in correspondence of the correlation for meridional velocity v.

Under the SSP585 scenario the relationships among the 4 variables are nearly invariant in MPI, but change in EC-Earth, especially in the low level wind field, followed by OLR. In EC-Earth the zonal component increases its correlation with the multivariate, temperature dominated representation, while the opposite is verified in v (see also App. C, Figure \ref{fig:mode_1_future}). The correlations found for the second and third Isomap components (see Figure \ref{fig:correlations_multivariate_vs_univariate}(b,d)) further elucidate how the relative role of the variables differ in the modeled versus observed dynamics. Isomap-2, for example, clearly indicates that both models significantly underestimate all correlations between univariate and multivariate representation but for v in historical times. This behavior changes in the future for EC-Earth, that displays even higher correlations than found in ERA5 for T, U and OLR, but it is unaltered in MPI.

\section{Local properties of the attractor} \label{sec:quantitative_results}

\subsection{Multivariate Fields} \label{sec:multivariate_metrics}

\begin{figure*}[tbhp]
\centering
\includegraphics[width=1\linewidth]{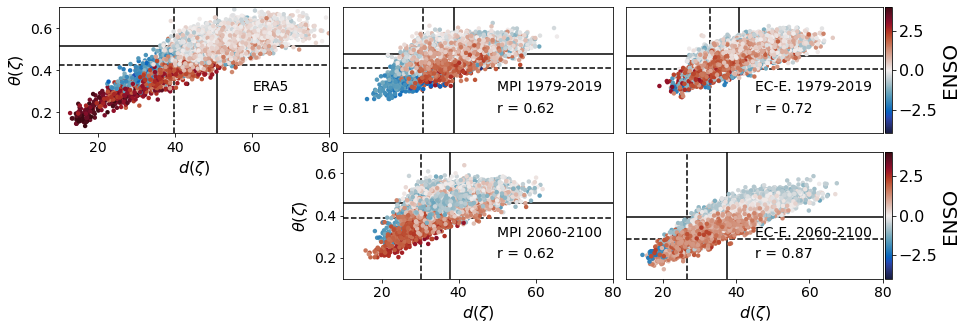}
\caption{Panel (a): Scatter plot of stability $\theta(\zeta)$ and local dimension $d(\zeta)$ for observations (ERA5) and the two models (MPI and EC-Earth). Each point represents a day in the period 1979-2019 (top panels) or 2060-2100 (bottom panels). Each point is colored according to its value in the first (standardized) Isomap component, quantifying here the ENSO index and dashed (solid) lines indicate the 0.1 (0.5) quantiles of $d(\zeta)$ and $\theta(\zeta)$. The correlation $r$ between each $d$ and $\theta$ is also reported. The computation spans the full 17,092-dimensional state space.}
\label{fig:theta_vs_d}
\end{figure*}

We quantify local properties of the high-dimensional flow through the local dimension  $d(\zeta)$ and persistence $\theta(\zeta)$ metrics introduced in Section \ref{sec:method}. Both metrics are calculated in the state space of the tropical Pacific.\\
The scatter plots of $d(\zeta)$ vs $\theta(\zeta)$ are shown in Figure \ref{fig:theta_vs_d}. Each point encodes a day in the $d(\zeta)$ vs $\theta(\zeta)$ space and is colored by its respective ENSO value, here defined by the first Isomap component (App. \ref{app:ENSO}, Figures \ref{fig:mode_1} and \ref{fig:mode_1_future}). In the observations the two metrics are strongly correlated ($r = 0.81$): days with lower local dimension are characterized by a large mean residence time and higher predictability (low values of $\theta(\zeta)$). Strong El Ni\~{n}os, and to a lesser extent, La Ni\~{n}as  are characterized by low $d(\zeta)$ and $\theta(\zeta)$ indicating that strong positive and negative ENSO events can be, to a first approximation, interpreted as unstable fixed points of the system. This analysis supports the nonlinear oscillator theoretical framework to explain ENSO dynamics \cite{AnSI0} and the asymmetry between El Ni\~{n}os and La Ni\~{n}as \cite{AnSI}. The asymmetry in the strength of positive and negative events is missed in both models, with the greater predictability of strong El Ni\~{n}os compared to strong La Ni\~{n}as being reversed in MPI in the historical period. Over the period 1979-2019, the correlation between $d(\zeta)$ and $\theta(\zeta)$ is $r = 0.62$ and $0.72$ for MPI and EC-Earth, respectively, therefore lower than in ERA5. Most importantly, the modeled values in the ($d(\zeta)$, $\theta(\zeta)$) diagram span a smaller region than in the reanalysis.\\

Indeed the region characterized by $d(\zeta) > 50 $ and $\theta(\zeta) > 0.5$  is very often explored in the reanalysis and is occupied by the majority of ENSO-neutral days, but is seldom visited by both MPI and EC-Earth. In other words the reanalysis captures higher dimensional  dynamics than the models. These differences are shared by all ensemble members analyzed. This analysis points to a large difference in predictability potential in both models compared to observations, with the models anomalies being far more predictable in neutral conditions. The  difference  is further  quantified  by  the  histograms of both metrics in Figure \ref{fig:theta_and_d_histogram} and by the four moments of each distributions summarized in Appendix \ref{app: moments}.\\

In the future, both models shift towards greater predictability potential, with lower values of $d(\zeta)$ and $\theta(\zeta)$ (see Figure \ref{fig:theta_and_d_histogram}). This shift to larger predictability in a warmer climate has been termed as “hammam effect” and first recognized in model simulations for the sea level pressure at the mid-latitudes \citep{hammam}, and in idealized aquaplanet simulations \cite{brunetti}. These changes are subtle in MPI and larger in EC-Earth, especially for persistence. In this model, the regular, periodic behavior of its future ENSO (see App. \ref{app:ENSO} Figure \ref{fig:mode_1_future}) causes the distribution of $\theta(\zeta)$ to shift to lower average and skewness values (Figure \ref{fig:theta_and_d_histogram} and Appendix \ref{app: moments}, Table \ref{tab:moments}). In MPI, El Ni\~{n}os  become  more  predictable  in the future and  have  a  lower instantaneous dimension than present ones, partially recovering the asymmetry bias found in the historical period.\\

\begin{figure}[tbhp]
\centering
\includegraphics[width=0.5\textwidth]{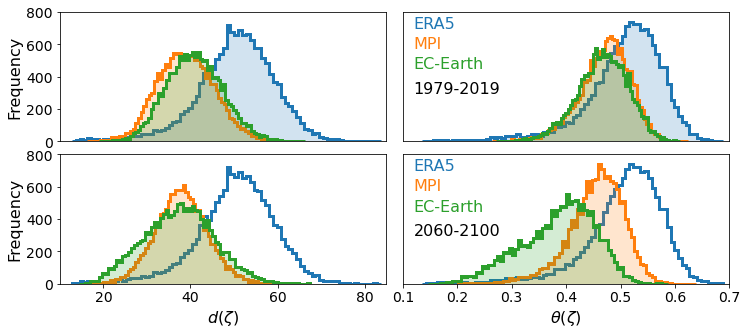}
\caption{Histograms of local dimension $d(\zeta)$ and inverse of persistency $\theta(\zeta)$ for ERA5, MPI and EC-Earth. Top panels: 1979-2019 period. Bottom panels: period 2060-2100 under the SSP585 scenario.}
\label{fig:theta_and_d_histogram}
\end{figure}

\begin{figure*}[tbhp]
\centering
\includegraphics[width = 0.8\linewidth]{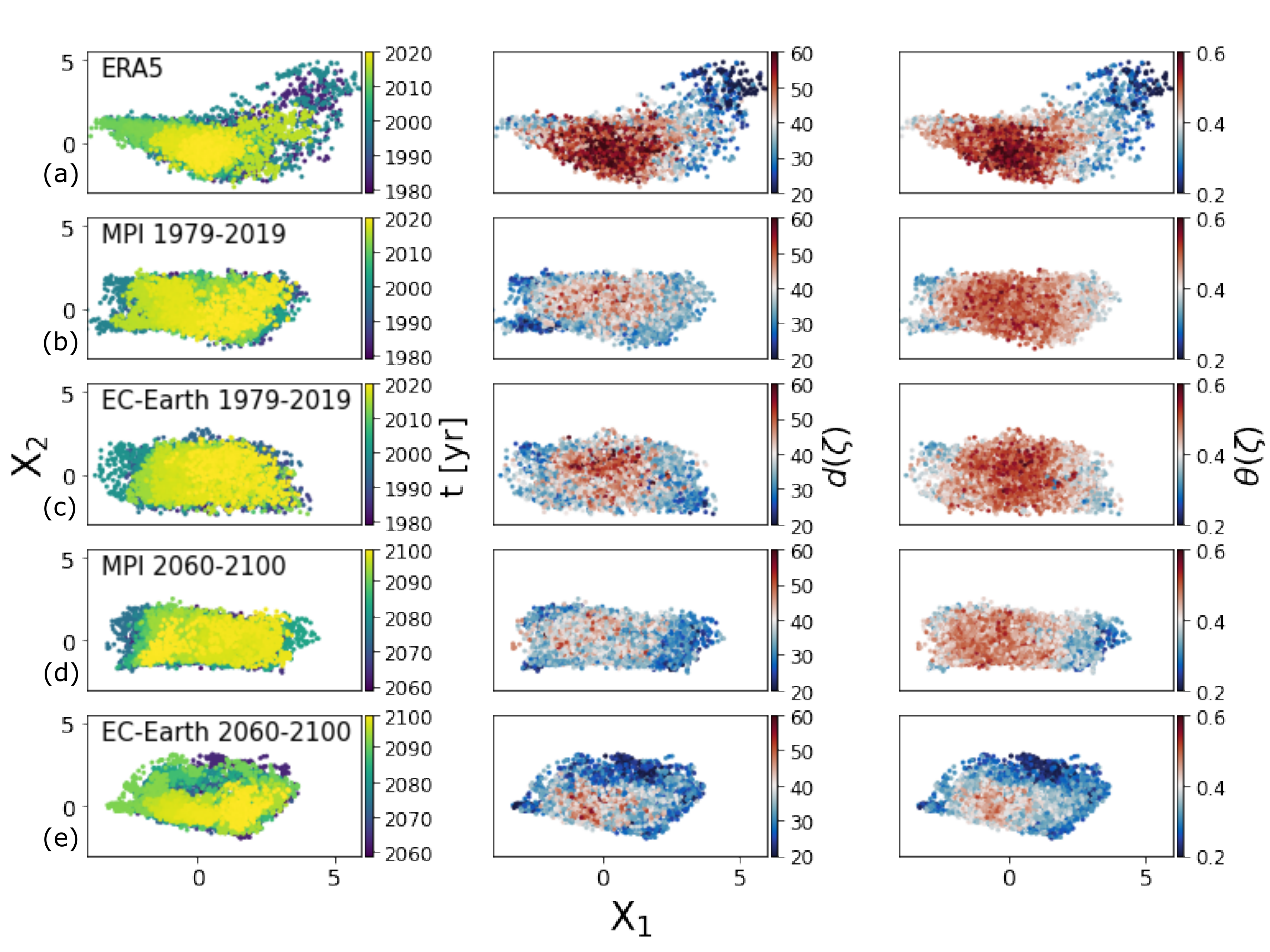}
\caption{In each row, from left to right: 2-dimensional, Isomap projections of the high-dimensional state space dynamics; Isomap projections with each point colored by its local dimension $d(\zeta)$; Isomap projections with each point colored by its local stability $\theta(\zeta)$. Row (a,b,c,d,e): ERA5 reanalysis, MPI (1979-2019), EC-Earth (1979-2019), MPI (2060-2100) and EC-Earth (2060-2100).}
\label{fig:theta_and_d_state_space}
\end{figure*}

Figure \ref{fig:theta_and_d_state_space} further visualizes the metrics and the differences between models and observations in the 2D (first and second component) Isomap projections. Both metrics, quantifying local geometry and stability, vary along the manifold, with low values at the outer borders (El Ni\~{n}o and La Ni\~{n}a regions) and high value close to the manifold center.\\

\subsection{Univariate fields and their scaling} \label{sec:scaling}

The analysis so far shows that distributions of local dimensions and persistence are biased towards lower values in the modelled climate. A lower average dimension $d$ implies that the models analyzed are inherently less complex; additionally, lower $\theta$ imply that anomalies in the Pacific, on average, persist longer than in observations.\\

Climate models do not solve all scales and therefore their intrinsic dimensionality (i.e., dimension of their inertial manifold) is expected to be lower than in  the reanalysis, as suggested in \cite{Dubrulle}; however, in spite of large differences in atmospheric horizontal resolution  ($\sim 100$km and $\sim 40$km for MPI and EC-Earth, respectively) and in parametrization schemes these two models have a nearly identical attractor dimensionality (see Appendix \ref{app: moments}, Table  \ref{tab:moments}).\\
We further investigate the source for this bias by exploring the sub-spaces defined by each univariate field. While lower dimensional, a model should properly capture the observed scaling among local dimension and persistence of each variable.\\

\begin{figure*}[tbhp]
\centering
\includegraphics[width = 1\textwidth]{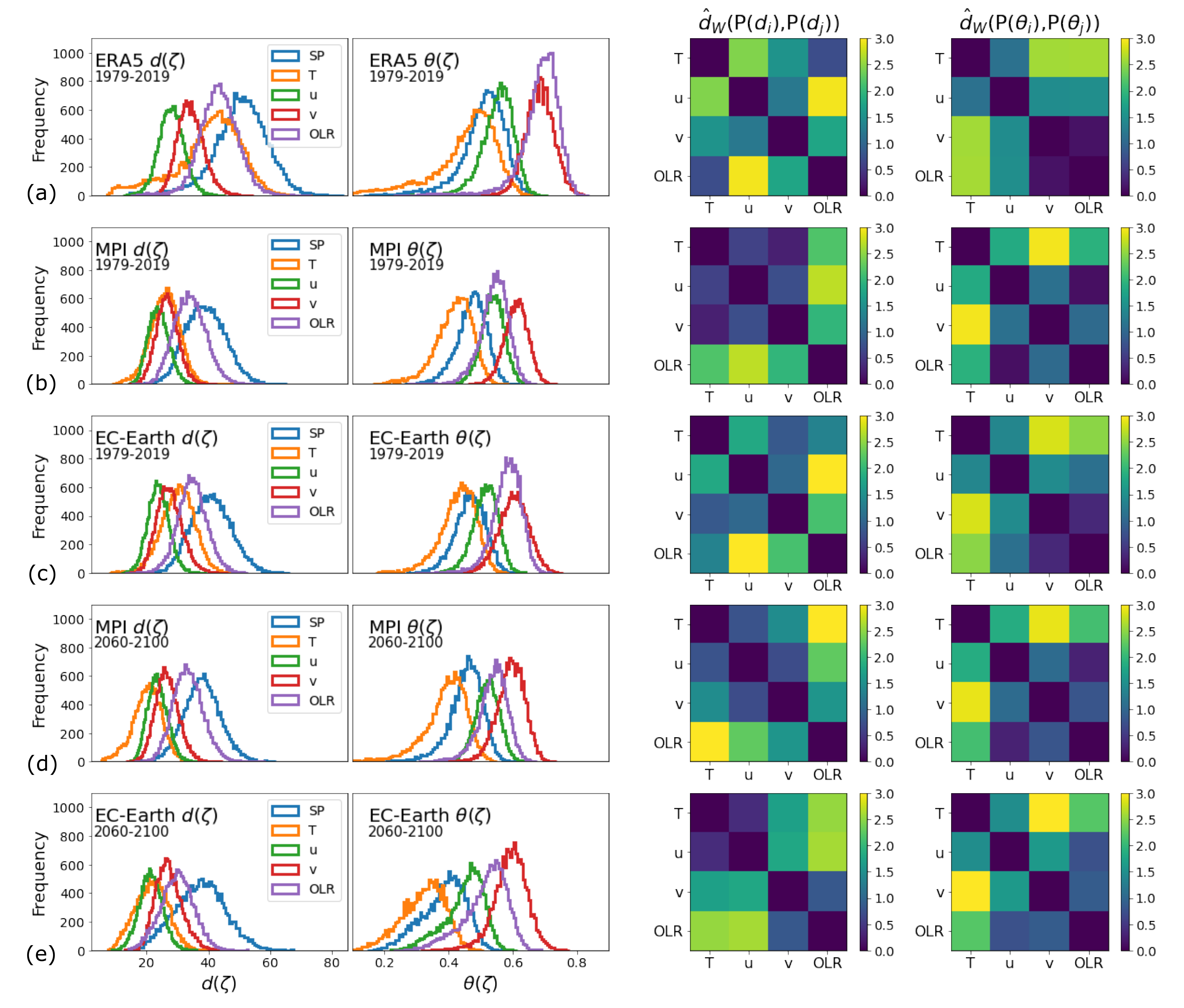}
\caption{First and second columns: Histograms of local dimension $d(\zeta)$ and inverse of persistence $\theta(\zeta)$ for the multivariate representation (referred to as state space, ``SP'') and univariate fields (i.e., T, u ,v and OLR). Third and fourth columns: Wasserstein distance between such histograms. For a given dataset, the variance among all distances is one. Row (a,b,c,d,e): ERA5 reanalysis, MPI (1979-2019), EC-Earth (1979-2019), MPI (2060-2100) and EC-Earth (2060-2100).}
\label{fig:theta_and_d_histogram_univariate}
\end{figure*}

We compute local dimension and persistence for each variable separately and quantify distances between ERA5 and models' distributions using the Wasserstein distance metric \cite{Villani}. Results are presented in terms of pairwise distance matrices. Crucially, we are interested in quantifying how distances between distributions scale among each other, rather than their absolute magnitude, therefore each pairwise distance matrix is further standardized by its total standard deviation. This analysis is shown in Figure \ref{fig:theta_and_d_histogram_univariate} for present and future periods.\\ 

In terms of local dimension, the distributions for ERA5 (Figure \ref{fig:theta_and_d_histogram_univariate}a, left panel) show that the manifold dimensionality of univariate fields is always smaller than the one embedded in a multivariate space, as to be expected.  The lowest dimensionality  characterizes the zonal velocity, followed by the meridional component, while similar average values (but not tails) are found for temperature  and  OLR. The strong skewness at low values in the temperature is linked to the most intense  ENSO events, as shown previously. In MPI the temperature  and velocity fields show similar values of local dimension with almost overlapping distributions between T and v. The OLR field has higher local dimensions. This lack of differentiation among variables is partially corrected in EC-Earth, but T and OLR distributions have different mean values.\\

By the end of the XXI century, the distributions of  local dimension for temperature shift towards lower values in both models, while changes are very limited for the other variables. The increase in intrinsic predictability under warming scenario around the Equator is therefore linked to changes in the surface temperature field alone.\\
Moving to persistence, distributions in ERA5 display low mean values of  $\theta$ for temperature and zonal velocity, and more limited predictability (higher $\theta$) for OLR and v. EC-Earth again reproduces the relative scaling and the relative distance among the distributions better than MPI. In the future, the distributions shift to lower values in EC-Earth, especially for temperature, while they remain nearly unchanged in MPI. Pairwise distance matrices, characterizing the degree of similarities among distributions, are then shown in the third column of Figure \ref{fig:theta_and_d_histogram_univariate}. They quantify from a dynamical perspective the relative ranking among MPI and EC-Earth noticed in \cite{Fasullo}. EC-Earth is indeed in better agreement with the reanalysis in terms of its representation of the relative contributions of each field to the multivariable distribution.

\clearpage

\section{Robustness of dynamical systems metrics: internal variability and resolution} \label{sec:robustness}

We conclude the presentation by analyzing the robustness of the dynamical system metrics to the internal variability of the (modeled) system and to the resolution chosen for the analysis.

\subsection{Internal variability} \label{sec: internal_variability}

\begin{figure}[tbhp]
\centering
\includegraphics[width = 1\linewidth]{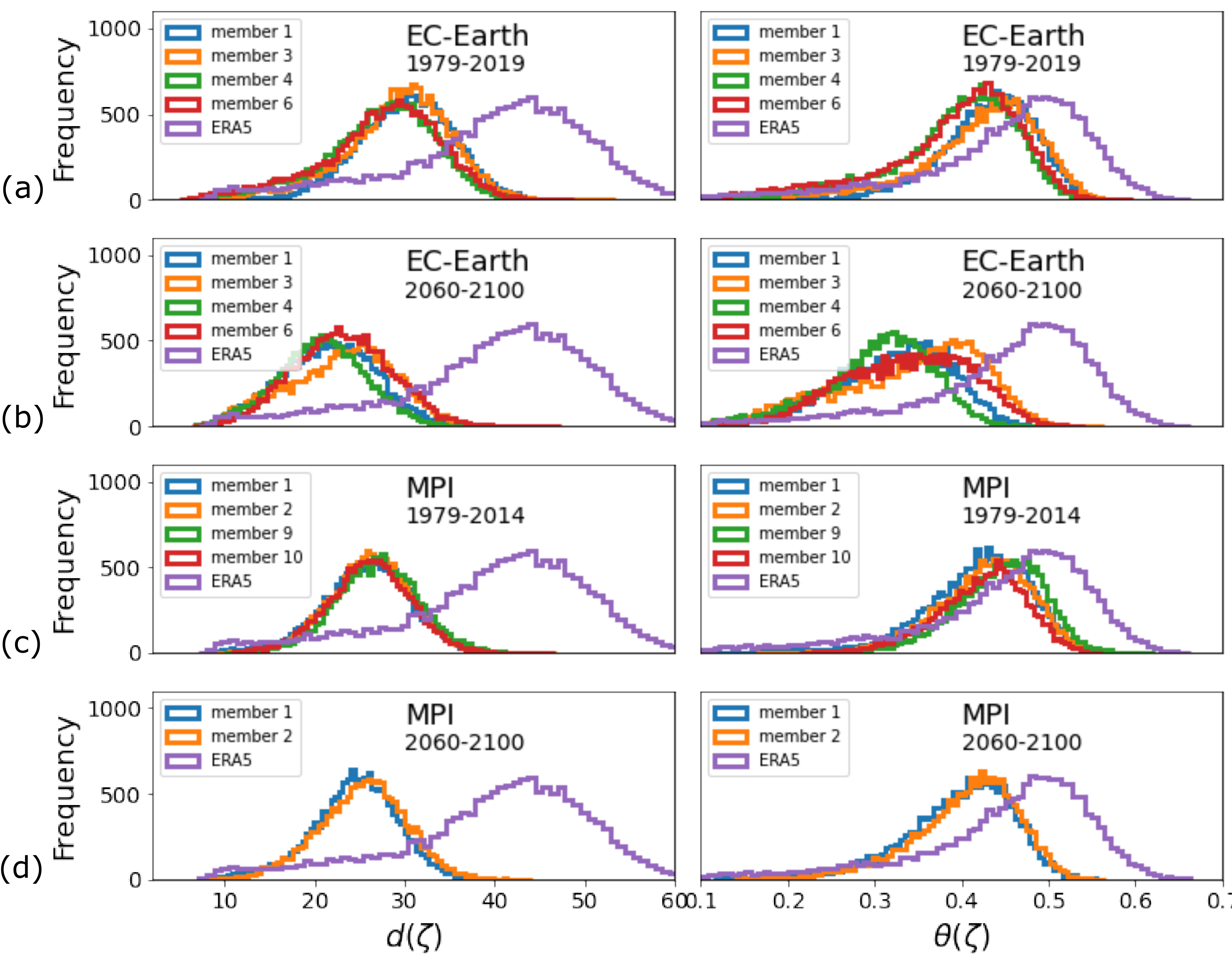}
\caption{Histograms of local dimension $d(\zeta)$ and inverse of persistence $\theta(\zeta)$ for the temperature field in EC-Earth and MPI. First two rows show results for EC-Earth. Last two rows show results for MPI. The analysis is performed for the temperature variable.}
\label{fig:internal_variability}
\end{figure}

We compute the local dimension and persistence metrics for 4 ensemble members in EC-Earth and in MPI. For the period 2060-2100 in MPI, we rely only on two members. For simplicity, this analysis focuses only on the temperature variable. Results are shown in Figure \ref{fig:internal_variability}. Chaotic trajectories of the same dynamical system are bounded to live on the same manifold and manifold properties are largely independent of the ensemble member, provided that we sample such object well enough. The similarities among members are further quantified by the first four moments of the distributions in Table \ref{tab:moments_internal_variability}. This analysis further show that the 40 years considered are sufficient to characterize the attractor. 

\begin{table}[tbhp]
\caption{\label{tab:moments_internal_variability} First four moments of the distributions of local dimensions and persistence. $\mu$, $\sigma$, $\gamma$, $\kappa$ are the mean, standard deviation, skewness and kurtosis of the different histograms, respectively. The subscript $d$ or $\theta$ indicates whether the analysis focuses on local dimension or persistence. Note: skewness and kurtosis for a normal distribution equal 0 and 3 respectively. ``MPI/EC-E. mi'' indicates the i-th member of the respective model ensemble.
}
\begin{ruledtabular}
\begin{tabular}{l*{8}{c}r}
&
\textrm{\large$\mu_d$}&
\textrm{\large$\sigma_d$}&
\textrm{\large$\gamma_d$}&
\textrm{\large$\kappa_d $}& &
\textrm{\large$\mu_\theta$}&
\textrm{\large$\sigma_\theta$}&
\textrm{\large$\gamma_\theta$}&
\textrm{\large$\kappa_\theta$}\\
\colrule
\tiny{1979-2019} &  &  &  &  &  & &  &  & \\
MPI m1 & 25.72 &  4.95 & -0.26 &  3.38 &  & 0.42 &  0.06 & -0.86 &  4.35\\
MPI m2 & 26.15 &  4.53 & -0.13 &  3.17 &  & 0.43 &  0.05 & -0.55 &  3.55\\
MPI m9 & 26.92 &  4.63 & -0.05 &  3.22 &  & 0.44 &  0.05 & -0.61 &  3.58\\
MPI m10 & 25.98 &  4.58 & -0.11 &  3.14 &  & 0.42 &  0.05 & -0.55 &  3.28\\
EC-E. m1 & 30.05 &  5.36 & -0.3 &  3.42 &  & 0.43 &  0.06 & -0.79 &  4.2\\
EC-E. m3 & 29.23 &  5.87 & -0.49 &  3.39 &  & 0.42 &  0.07 & -0.95 &  3.89\\
EC-E. m4 & 27.22 &  6.21 & -0.55 &  3.32 &  & 0.39 &  0.07 & -0.97 &  3.86\\
EC-E. m6 & 27.34 &  6.68 & -0.58 &  3.27 &  & 0.39 &  0.08 & -1.05 &  3.91\\
\tiny{2060-2100} &  &  &  &  &  & &  &  & \\
MPI m1 & 24.39 &  4.83 & -0.37 &  3.43 &  & 0.4 &  0.06 & -0.84 &  3.99\\
MPI m2 & 25.26 &  4.87 & -0.2 &  3.09 &  & 0.4 &  0.06 & -0.72 &  3.55\\
EC-E. m1 & 21.54 &  5.16 & 0.00 & 2.67 &  & 0.32 &  0.07 & -0.46 &  2.84\\
EC-E. m3 & 22.91 &  5.87 & -0.21 &  2.50 &  & 0.34 & 0.08 & -0.53 & 2.54\\
EC-E. m4 & 20.76 &  4.89 & -0.10 &  2.90 &  & 0.3 &  0.06 & -0.58 & 3.18\\
EC-E. m6 & 23.01 &  5.43 & 0.05 &  2.86 &  & 0.34 &  0.07 & -0.26 &  2.57\\
\end{tabular}
\end{ruledtabular}
\end{table}

\subsection{Resolution dependence} \label{sec:resolution_dependence}

To compute the dynamical system metrics we first define an observable as $g(\mathbf{X}(t),\zeta)=-\log(\delta(\mathbf{X}(t),\zeta))$, where $\mathbf{X}(t)\in\mathbb{R}^{\textit{T,N}}$ represents the high-dimensional trajectory in a state space of dimensionality $N$ (Section \ref{sec:method}). Here the function $\delta(\mathbf{X}(t),\zeta)$ represents the Euclidean distance between a state space point $\zeta$ and the trajectory $\mathbf{X}(t)$. Computations of distances in high dimensions are affected by the known ``curse of dimensionality'' \cite{pons}, and it is therefore important to check how results differ with the fields' resolution.\\
Here we focus on the $d$ and $\theta$ metrics in the multivariate and univariate representations at two different resolutions.

For ERA5 we consider:

\begin{itemize}
    \item  the upscaled resolution of 1$^o$ in latitude and 1.5$^o$ in longitude adopted in most of our analysis. This implies a multivariate embedding in a $N = 17,092$ dimensional state space and a univariate embedding in a $N = 4,273$ dimensional state space.
    \item A higher resolution with 0.5$^o$ in both latitude and longitude. This implies a multivariate embedding in a $N = 100,956$ state space and a univariate embedding in a $N = 25,239$ dimensional state space.
\end{itemize}

\begin{figure}[tbhp]
\centering
\includegraphics[width=1.\linewidth]{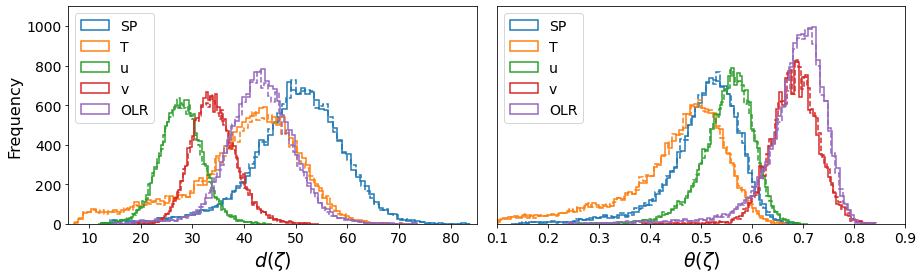}
\caption{Histograms of local dimension $d(\zeta)$ and inverse of persistence $\theta(\zeta)$ for the multivariate representation (referred to as state space, ``SP'') and univariate fields (i.e., T, u ,v and OLR) in ERA5. Dashed lines: univariate and SP representations living in  $N = 25,239$ and $N = 100,956$ respectively. Solid lines: univariate and SP representations living in  $N = 4,273$ and $N = 17,092$ respectively.}
\label{fig:robustness_res_era5}
\end{figure}

Results are independent of resolution, given that the system has a large scale imprinting which is captured in both cases (as shown in Figure \ref{fig:robustness_res_era5}).\\

\begin{figure}[tbhp]
\centering
\includegraphics[width=1.\linewidth]{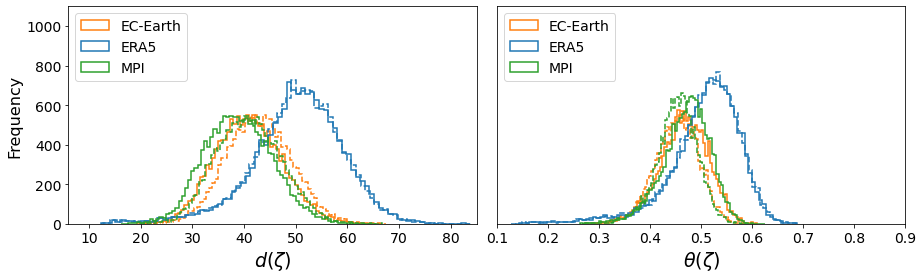}
\caption{Histograms of local dimension $d(\zeta)$ and inverse of persistence $\theta(\zeta)$ for the multivariate representation in the two models. Dashed lines: high resolution case, with  ERA5, EC-Earth and MPI state space dimensionality $N$ equal to $100,956$; $52,7809$ and $28,344$ respectively. Solid lines: upscaled (same) resolution $N = 17,092$ for ERA5 and both models.}
\label{fig:robustness_sp_models}
\end{figure}

The same is verified in the models, where we compare the multivariate state space dynamics for the upscaled resolution of 1$^o$ in latitude and 1.5$^o$ in longitude and the original model resolution, for which $N = 52,780$ for EC-Earth, and $N = 28,344$ for MPI. The analysis is shown in Figure \ref{fig:robustness_sp_models}. Differences among models with original or upscaled resolutions are minimal. For reference, we also plot the ERA5 results for the low and high resolution case.

\clearpage

\section{Discussion and future work} \label{sec:discussion}

This work introduces a powerful framework stemming from dynamical system theory to investigate climate variability and account for its spatiotemporal and multivariable  dependency. The methodology is based on the assumption that the high-dimensional trajectory  of  the  climate system lives on a lower dimensional manifold  \cite{Predrag,GhilLucarini}. 

Characterizing the topology and quantifying the geometrical properties of the climate attractor, alongside dynamical properties of the trajectories on the manifold, offers a much needed, robust framework for dimensionality reduction in climate studies.
Here we considered the tropical Pacific and four variables that are key to its variability, and explored the high-dimensional system’s dynamics in an observational dataset, ERA5, and in realizations of two state-of-the-art climate models, MPI and EC-Earth. The analysis studied the high-dimensional tropical Pacific dynamics through manifold learning algorithms and dynamical system metrics. This provided a first estimate of the dimensionality of the tropical Pacific manifold, which is $\sim$50 for ERA5, and around 39 in the models.\\

A first, important result is that in ERA5 the nonlinear algorithm always shows a faster saturation of the residual variance compared to PCA, so that the dynamics can be projected onto  fewer  dimensional  components. Independent of the dataset, the  dynamics  lives  on  a torus if the seasonal cycle is included. While the torus is topologically similar among datasets, large excursions in  correspondence  of  El  Ni\~{n}o  events  are  found  in  the observations but are absent in the models. Furthermore, the spatial signature of the seasonal cycle, described by the first Isomap component, is biased in both models.
By analyzing the anomalies, we showed that the two  models have qualitative and quantitative similarities in the geometrical properties of their manifold,  despite different resolutions and different choices in the representation of small scale, unresolved processes. The PCA residual variance   shares similarities among observational and modeled datasets, while the Isomap residual variance differs from that of PCA in the reanalysis while remaining similar in both models. This implies that both MPI and EC-Earth struggle in capturing the nonlinear topological characteristics of the observed manifold, and they do so in a similar way. Differences between the observed and modeled ENSO, on the other hand, are model dependent and larger in MPI for the historical period.

We stress that a key aspect of this work is the inclusion  of multiple variables for a more comprehensive and robust quantification of the system’s dynamics and feedbacks. The comparison between multivariate and univariate properties of the attractors quantifies the relative contribution of each field and allows for evaluating how this contribution may change over time. In the tropical Pacific the temperature field dominates the variance in both models and reanalysis, but the correlation between each field and the embedded trajectory differs in the models compared to the reanalysis and evolves differently in the two models in the scenario considered. It is also interesting to see how the relative role of the variables differ among the models, while contributing to a similar attractor in the historical periods, when models can be tuned through parameter choices towards the observations.\\ 

This work opens the way to evaluating the attractor trajectories over time in models and comparing them to the observed one in the past 40 years, to better constrain climate sensitivity and the evolution of climate feedbacks, both imperative to predict the likelihood of tipping points in the system.

By adopting the local geometry and persistence metrics to characterize the attractor’s properties, we neatly visualized the day-to-day predictability potential during ENSO  events,  the  El  Ni\~{n}o/La  Ni\~{n}a  asymmetry, and model biases with regard to both aspects. Differences between the attractor in the models and reanalysis are not limited to strong ENSO events.  Indeed, the region with both high $d(\zeta)$ and $\theta(\zeta)$, which is the most explored in the reanalysis and corresponds to ENSO-neutral days, is very  seldom  occupied  by  the  models,  implying  that the representation of locally unstable motions and their influence on large-scale climate dynamics continues to elude current climate models, and such elusion is not amended in EC-Earth, despite being run at higher resolution than MPI. 

These results point to topological (global, in state space) and geometrical (local, in state space) differences between observationally-based data and climate model outputs, which can be evaluated considering one simulation, without the need for computationally expensive ensembles.  The local scale chaoticity of the climate modeled system remains underestimated in both models, notwithstanding their different resolutions. Furthermore, the relationships among variables, which set their contribution to the global attractor and are fundamental to the evolution of the climate systems, are misrepresented, in different ways, in both models, and more so in MPI. The quantification of these relationships is a key, novel outcome of this work.\\ 

The framework we propose can be adopted to evaluate in a straightforward, robust way the impact of parameterizations on the (modeled) climate manifold and therefore assessing their impact on the large-scale dynamics. Most importantly, the analysis sets the stage for manifold learning approaches to climate modelling and climate prediction based not only on small-scale process understanding (machine learning application to subgrid scale parameterizations), but also on the characterization of the global climate system topology and the relations among variables (relational probabilistic models \cite{Raedt}). In the future, novel approaches stemming from data-driven dynamics and control (e.g., \cite{Loiseau,Champion}) could be adopted to learn reduced-order models governing the evolution of the effective degrees of freedom of the system, therefore providing a useful alternative to the traditional partial differential equation (PDE) approaches adopted in climate science.

\section*{Acknowledgments}

We thank Predrag Cvitanovi\'{c} and the \href{http://chaosbook.org/}{ChaosBook} team who influenced many of the ideas in this work. We thank Yoann Robin for the online implementation of dynamical system metrics and Jost von Hardenberg for clarifications on the EC-Earth3, CMIP6 runs.  F.F. acknowledges helpful discussions on the topic with Sebastián Ortega, Sylvia Sullivan and Ilias Fountalis. The authors were supported by the Department of Energy, Regional and Global Model Analysis (RGMA) Program, Grant N.: 0000253789. The revision process benefitted from the KITP Program ``Machine Learning and the Physics of Climate'' supported by the National Science Foundation under Grant No. NSF PHY-1748958. Furthermore, we especially thank the editorial team and two anonymous reviewers for their insightful comments that have greatly improved this paper.

\section*{Code availability}

A Python implementation of dynamical system metrics is freely available at https://github.com/yrobink/CDSK/tree/master/python. 
For the PCA and Isomap algorithm we adopted the implementation in the Scikit-learn library \cite{scikit-learn}. A Github repository with examples and updates on current work can be found here https://github.com/FabriFalasca/climate-and-dynamical-systems .

\appendix

\section{Is the manifold nonlinear? Euclidean vs geodesic distances} \label{app:nonlinearity}

To prove that the tropical Pacific manifold is indeed nonlinear we compute the Euclidean and geodesic distances between all pairs of points $A, B \in \mathbb{R}^{\textit{N}}$, where $N$ is the dimensionality of the state space. Under the assumption that the high-dimensional data live on a low-dimensional object $\mathcal{M} \in \mathbb{R}^{\textit{d}}$ (with $d<<N$), we face  two possibilities:
\begin{itemize}
    \item the manifold $\mathcal{M}$ is linear. In this case the Euclidean distances between each pair of points $A$ and $B$ have to be equal to their geodesic distances {\it along} the manifold;
    \item the manifold $\mathcal{M}$ is nonlinear (i.e., $\mathcal{M}$ is a curved object). In this case the geodesic distances {\it along} the manifold between each pair of points $A$ and $B$ are {\it always} greater than their Euclidean distances. This follows from the simple fact that the Euclidean distance is the shortest distance between two points.
\end{itemize}

Computing the geodesic distance is an important step of the Isomap algorithm \cite{ISOMAP}.\\
To compute the geodesic distances {\it along} the manifold we assume that, while nonlinear, the manifold is {\it locally} flat around a radius of $K$ points. It follows that the distances between each pair of points inside their $K$-neighborhood is the Euclidean distance $\delta_{E}$. We can therefore construct a weighted graph such that (a) each point $i$ and $j$ is connected if inside their $K$-neighborhood and (b) their connection is weighted by the distance $\delta_{E}(i,j)$. The geodesic distance $\delta_{G}(i,j)$ is then the shortest path between each pair $(i,j)$. For the shortest path computation we adopted the Floyd–Warshall algorithm  \cite{introToAlg}.\\

\begin{figure}[tbhp]
\centering
\includegraphics[width=0.5\textwidth]{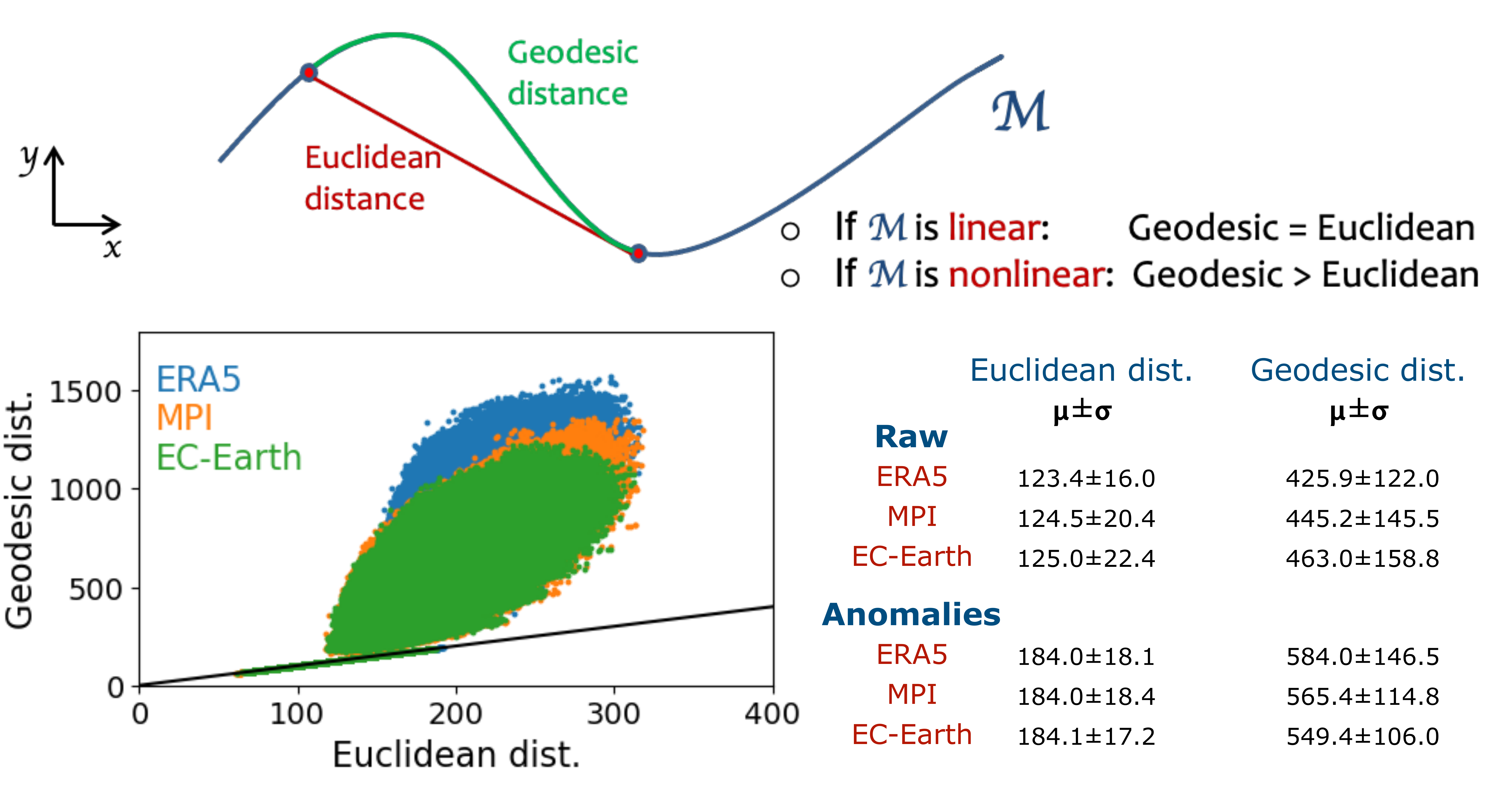}
\caption{Top panel: schematic to explain differences between geodesic ({\it along} the manifold) and Euclidean distances. In the schematic the state space is two-dimensional and the manifold $\mathcal{M}$ is simply a line. Bottom panel: geodesic ({\it along} the manifold) and Euclidean distances between each pair of points in each datasets for anomalies (no seasonality and no trends) in the period 1979-2019. Points are above the diagonal, therefore quantifying the intrinsic nonlinearity of the low-dimensional manifold. Distances are computed in the full 17,092-dimensional state space. Mean $\mu$ and standard deviation $\sigma$ for all three datasets are reported in the case of raw and anomalies. Raw data include trends.}
\label{fig:distances}
\end{figure}

We show the result in Figure \ref{fig:distances} for the period 1979-2019. Euclidean and geodesic distances are on the $x$ and $y$ axis respectively. As points are above the diagonal we can conclude that the manifold is indeed nonlinear for all three datasets analyzed. In this paper $K = 10$ days. 
Note of caution: the raw data include trends while the dataset with anomalies has been linearly detrended. 

\section{Seasonal cycle} \label{app:seasonal_cycle}

\begin{figure}[tbhp]
\centering 
\includegraphics[width=0.5\textwidth]{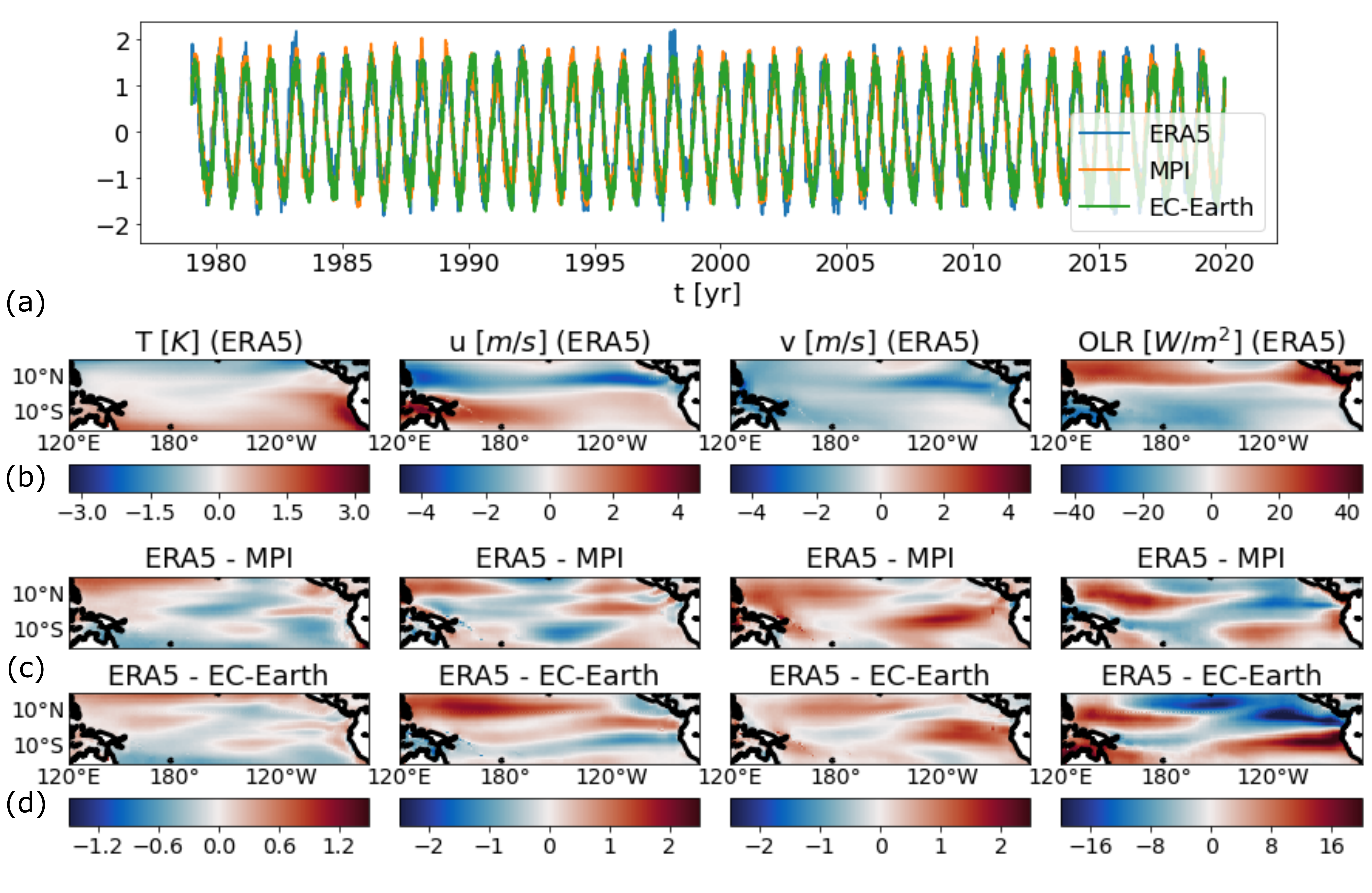}
\caption{Panel (a): first (standardized) Isomap component for  observations (ERA5) and the two CMIP6 models (MPI and EC-Earth). Each component is standardized to unit variance. Correlations with the first Principal component is higher than 0.98 independent of the dataset. Panel (b) top row: linear regression of the first Isomap component onto ERA5. Panel (b) bottom rows: differences between spatial projections of ERA5 and the two models. Note that the projections are spatial signatures of a single eigenvector and not four.}
\label{fig:seasonality}
\end{figure}

In Figure \ref{fig:seasonality}a, we consider the first Isomap component as described in Section ``A  first  test:   Mean  State  and  Seasonal  Cycle.''. This component represents the seasonal cycle. The Pearson correlation across datasets is always higher than 0.95, indicating that the temporal characteristic of the seasonal cycle are well captured by the two models. Independent of the dataset we find correlations higher than 0.98  between the first components of Isomap and PCA. Similarities across datasets in this first component do not imply similar spatial projections (i.e., the seasonal component of a model may be linked to biased regional processes even if highly correlated to observations). Spatial projections are visualized as the linear regression of an Isomap component onto each variable, and they therefore capture differences in variance. In Figure \ref{fig:seasonality}b we show the spatial signature of the first Isomap component for ERA5 and the differences between MPI and EC-Earth. All time series in this analysis are standardized to unit variance. Future studies will focus on differences in the signals amplitudes.\\ 

\section{ENSO: projections, spectral properties and spatial signatures} \label{app:ENSO}

\begin{figure*}[tbhp]
\centering
\includegraphics[width=0.7\linewidth]{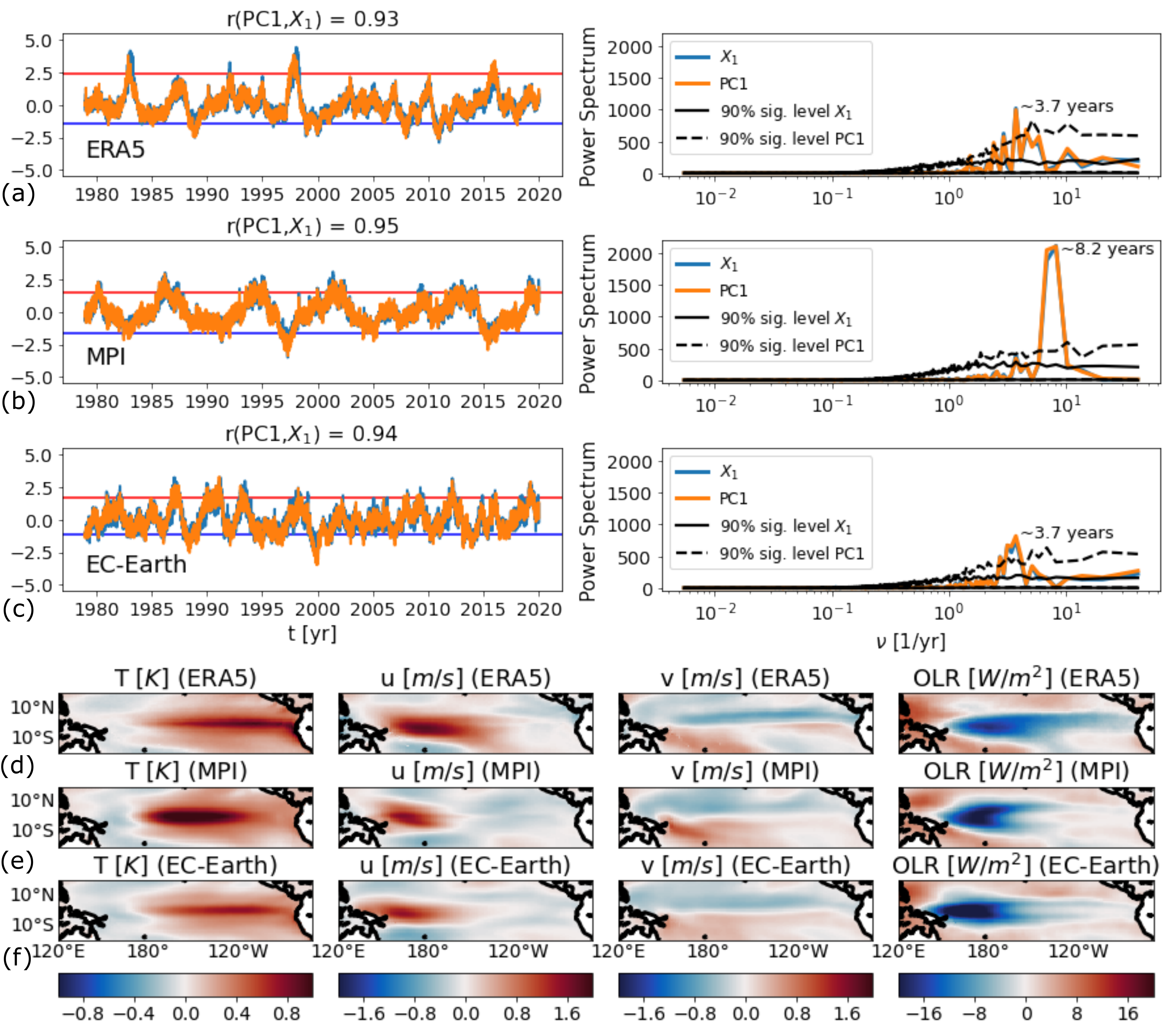}
\caption{Panel (a-c). Left column: First (standardized) Isomap ($X_{1}$) and Principal Component (PC1) for observations (ERA5) and the two CMIP6 models (MPI and EC-Earth). Red (blue) lines represent the median of positive (negative) values in  the 0.1 quantile (10-th percentile) of the joint PDF of $d$ and $\theta$ (see Figure \ref{fig:theta_vs_d}). Right column: correspondent Fourier spectra. The spectral significance has been tested under the null hypothesis of red noise \citep{Imkeller, Henk1,SmithRedNoise}. Panel (d-f): projection of the first (embedded) Isomap component on the various datasets.}
\label{fig:mode_1}
\end{figure*}

\begin{figure*}[tbhp]
\centering
\includegraphics[width=0.7\linewidth]{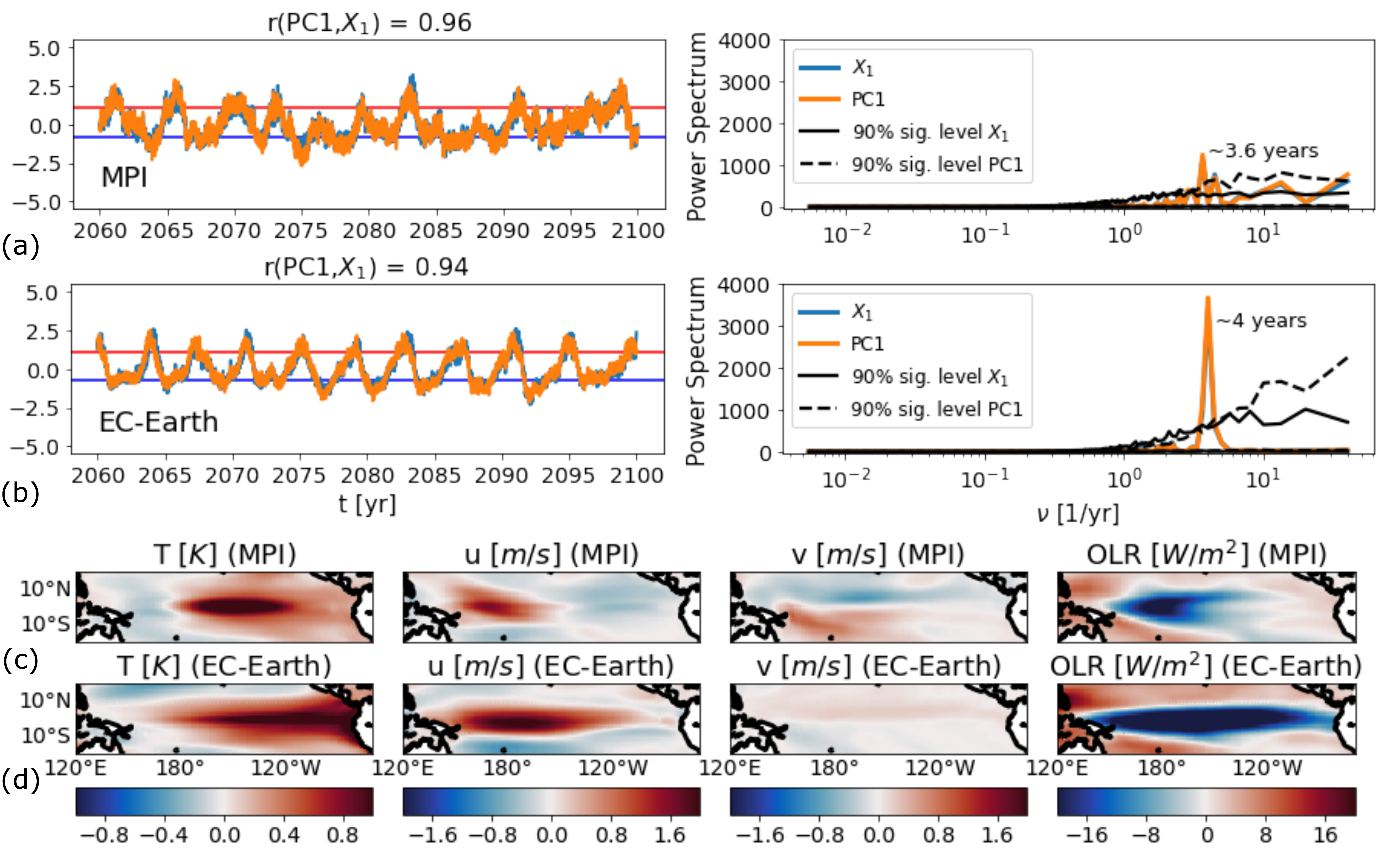}
\caption{Panel (a-b). Left column: First (standardized) Isomap ($X_{1}$) and Principal Component (PC1) for the two CMIP6 models (MPI and EC-Earth). Red (blue) lines represent the median of positive (negative) values in  the 0.1 quantile (10-th percentile) of the joint PDF of $d$ and $\theta$ (see Figure \ref{fig:theta_vs_d}). Right column: Fourier spectra. The spectral significance has been tested under the null hypothesis of red noise \citep{Imkeller, Henk1,SmithRedNoise}. Panel (c-d): projection of the first (embedded) Isomap component on the modeled datasets. The period analyzed is 2060-2100 under the SSP585 scenario.}
\label{fig:mode_1_future}
\end{figure*}

In Figures \ref{fig:mode_1}(a-c) and \ref{fig:mode_1_future}(a-b) we show the first Isomap and PCA components and their Fourier spectra in the period 1979-2019 and 2060-2100 respectively. Both components have been standardized to unit variance for comparison. Figures \ref{fig:mode_1}(d-f) and \ref{fig:mode_1_future}(c-d) show the spatial signature of the Isomap component on the four fields analyzed.\\


\clearpage

\section{Multivariate or univariate?} \label{app:1_variable}

\begin{figure}[tbhp]
\centering
\includegraphics[width=0.4\textwidth]{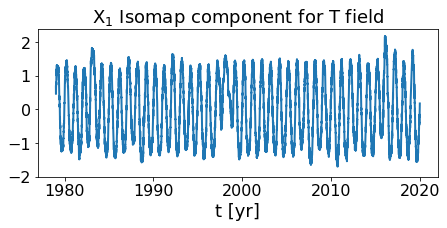}
\caption{First Isomap component of temperature anomalies in the ERA5 dataset.}
\label{fig:multivariate_vs_univariate_era5_onlyT}
\end{figure}

\begin{figure}[tbhp]
\centering
\includegraphics[width=1\linewidth]{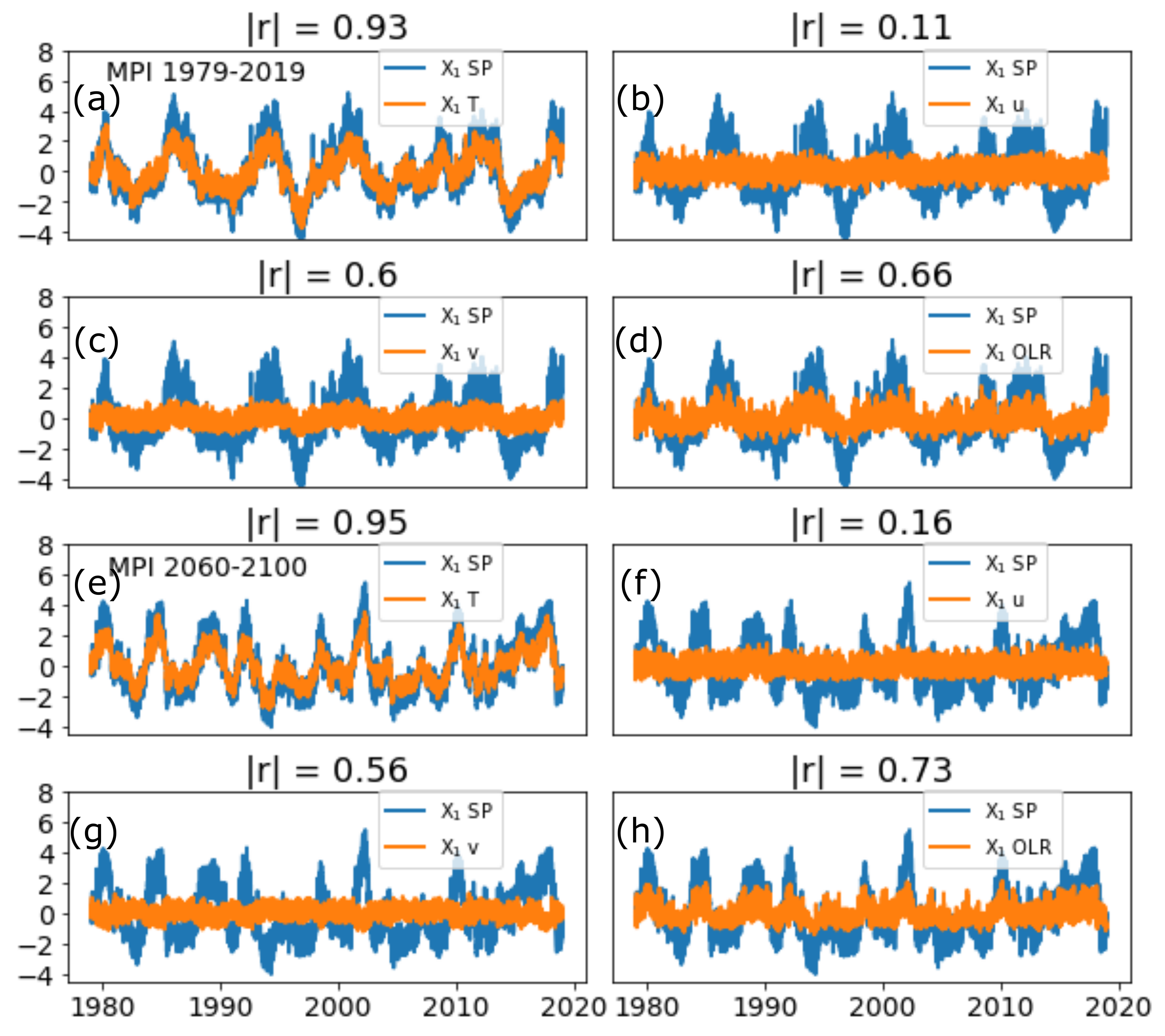}
\caption{First Isomap ($X_{1}$) component for anomalies in periods 1979-2019 and 2060-2100 in the MPI model. In each case components have been standardized so that the \textit{total} variance is equal to 1. Projections in the multivariate case are shown in blue and labelled as ``SP'' (state space). Projections for each, univariate field are shown in red. Atop of each plot we report the correlation coefficient (in absolute value) between the projections in the multivariate and univariate cases.}
\label{fig:models_multivariate_vs_univariate_MPI}
\end{figure}

\begin{figure}[tbhp]
\centering
\includegraphics[width=1\linewidth]{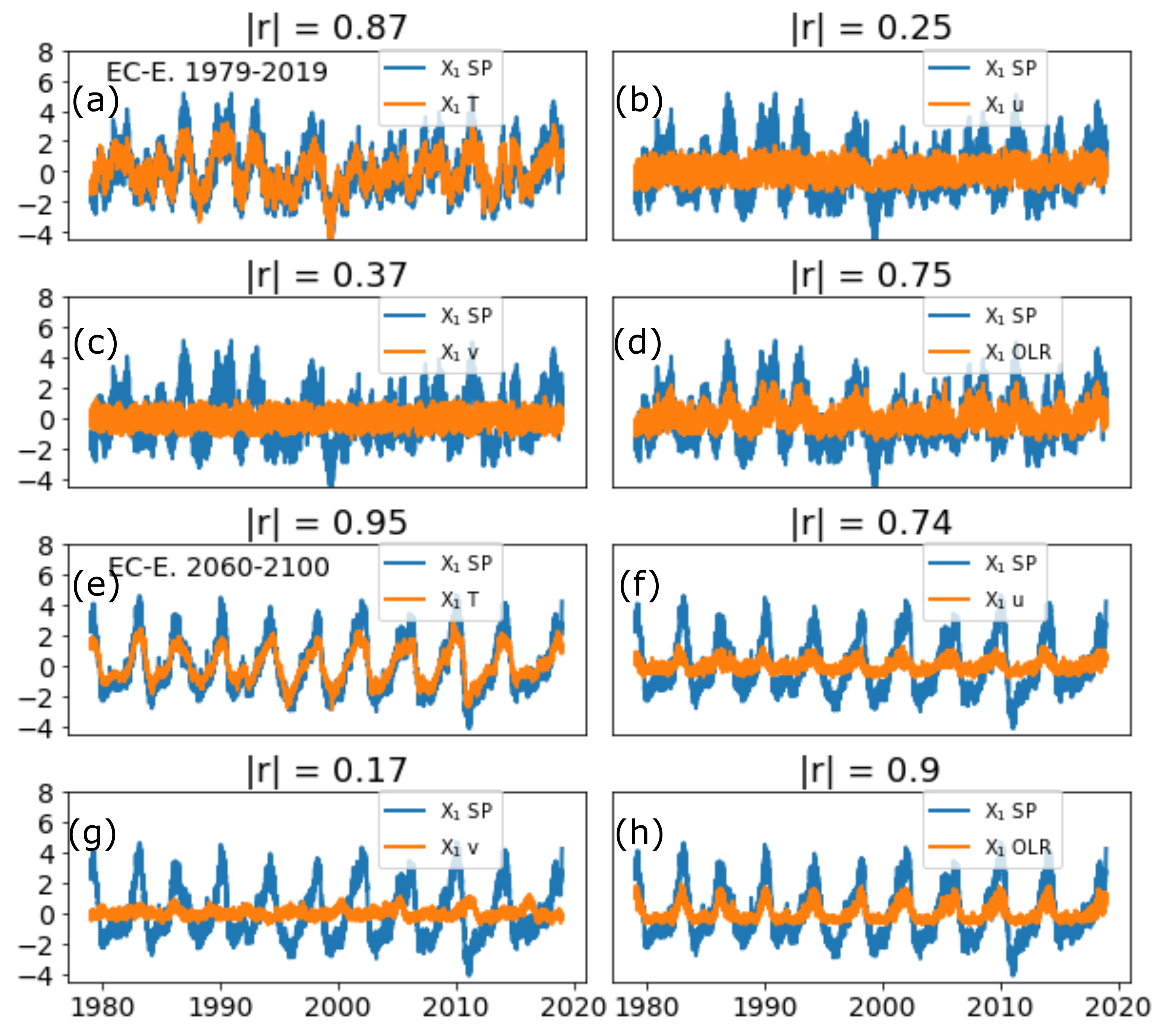}
\caption{First Isomap ($X_{1}$) component for anomalies in periods 1979-2019 and 2060-2100 in the EC-Earth model. In each case components have been standardized so that the \textit{total} variance is equal to 1. Projections in the multivariate case are shown in blue and labelled as ``SP'' (state space). Projections for each, univariate field are shown in red. Atop of each plot we report the correlation coefficient (in absolute value) between the projections in the multivariate and univariate cases.}
\label{fig:models_multivariate_vs_univariate_EC_Earth}
\end{figure}

Figure \ref{fig:multivariate_vs_univariate_era5_onlyT} show the first Isomap for the temperature field in ERA5. The seasonal cycle deviates from its canonical behavior in correspondence of large El Ni\~{n}o events.
In Figures \ref{fig:models_multivariate_vs_univariate_MPI} and \ref{fig:models_multivariate_vs_univariate_EC_Earth} we compare the first Isomap component obtained in the multivariate representation with the univariate case, for MPI and EC-Earth respectively.


\clearpage

\section{First four moments of the distributions of local dimensions and persistence} \label{app: moments}

In Table \ref{tab:moments} we show the first four moments of the distributions in Figure \ref{fig:theta_and_d_histogram}. The forth moment (i.e., kurtosis) in ERA5  highlights the presence of larger values in the tails of both the $d$ and $\theta$ distributions.

\begin{table}[tbhp]
\caption{\label{tab:moments} First four moments of the distributions of local dimensions and persistence. $\mu$, $\sigma$, $\gamma$, $\kappa$ are the mean, standard deviation, skewness and kurtosis of the PDFs, respectively. The subscript $d$ or $\theta$ indicates the different PDFs. Note: skewness and kurtosis for a normal distribution equal 0 and 3 respectively.
}
\begin{ruledtabular}
\begin{tabular}{l*{8}{c}r}
&
\textrm{\large$\mu_d$}&
\textrm{\large$\sigma_d$}&
\textrm{\large$\gamma_d$}&
\textrm{\large$\kappa_d $}& &
\textrm{\large$\mu_\theta$}&
\textrm{\large$\sigma_\theta$}&
\textrm{\large$\gamma_\theta$}&
\textrm{\large$\kappa_\theta$}\\
\colrule
\tiny{1979-2019} &  &  &  &  &  & &  &  & \\
ERA5 & 50.38 & 8.85 & -0.59 & 4.21 &  & 0.51 & 0.07 & -1.28 & 5.92\\
MPI & 38.92 & 6.45 & 0.11 & 2.88 &  & 0.47 & 0.05 & -0.58 & 3.65\\
EC-E. & 40.91 & 6.41 & 0.15 & 2.90 &  & 0.47 & 0.05 & -0.32 & 3.10\\
\tiny{2060-2100} &  &  &  &  &  & &  &  & \\
MPI & 37.89 & 5.98 & 0.06 & 3.19 &  & 0.45 & 0.05 & -0.82 & 4.42\\
EC-E. & 37.33 & 8.02 & 0.07 & 2.76 &  & 0.38 & 0.07 & -0.46 & 2.75\\
\end{tabular}
\end{ruledtabular}
\end{table}

\section{Dependence on q} \label{app:param_q}

\begin{figure}[tbhp]
\centering
\includegraphics[width=1.\linewidth]{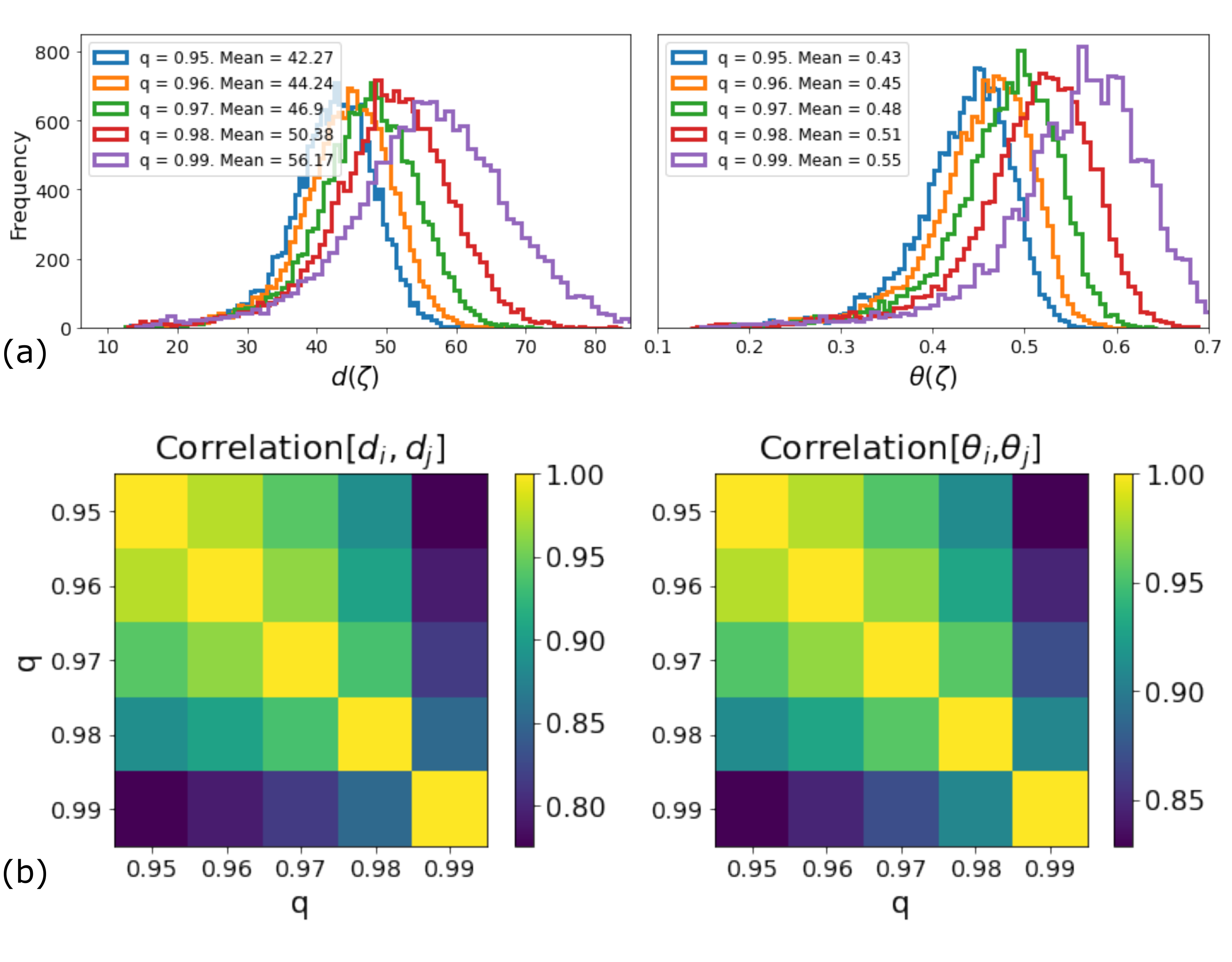}
\caption{Row (a): histograms of local dimension $d$ and persistence $\theta$ identified using different $q$. Row (b): pairwise correlation matrix of $d$ and $\theta$ identified using different $q$. The analysis is performed on the multivariate state space representation.}
\label{fig:robustness_q}
\end{figure}

We tested the robustness of the dynamical system metrics under the choice of threshold $q$ (see Section \ref{sec:method}) for the multivariate case. Robustness is evaluated for the range $q \in [0.95,0.99]$. We propose two analyses: first we compute histograms of both $d$ and $\theta$ and second we compare the temporal variability of the two metrics by looking at their correlation. The analysis is shown in Figure \ref{fig:robustness_q}. We first look at the average manifold dimension. Values vary from $\sim 42$ to $\sim 56$. This is quite a small range as we started from a noisy, $\sim 17000$ dimensional dynamical system. In terms of variability (see Figure \ref{fig:robustness_q}(b)) we see large correlations independently of the $d$ or $\theta$ variable. The largest outlier is $q = 0.99$.\\

We choose the value of $q = 0.98$ as it gives similar results with the $q = 0.95,0.96,0.97$ thresholds while still being a very high quantile (preferred as we are quantifying statistics of extremes).

\clearpage

\bibliography{references}

\end{document}